\documentstyle[preprint,aps,eqsecnum]{revtex}

\newcommand{\be}{\begin{equation}}
\newcommand{\ee}{\end{equation}}
\newcommand{\bea}{\begin{eqnarray}}
\newcommand{\eea}{\end{eqnarray}}
\newcommand{\pint}{\int \frac{d^3p}{(2\pi)^3}}
\newcommand{\eint}{\int_m^{\Lambda_E}}
\newcommand{\limit}{\lim_{\epsilon\to 0}}
\newcommand{\sgn}{{\rm sgn}\/}
\newcommand{\Vp}{\, {\cal P} \! \! \! \int}
\newcommand{\fspace}{\ }
\newcommand{\MeV}{{\rm MeV}}
\newcommand{\Punkt}{\quad .}
\newcommand{\Komma}{\quad ,}

\begin{document}
\draft

\preprint{HD--TVP--95--13}

\title{\vspace*{3cm} One Loop Integrals at \\ Finite Temperature and Density}

\author{P.~Rehberg\thanks{E-Mail:
        {\tt Peter@Frodo.TPhys.Uni-Heidelberg.DE}}
        and S.~P.~Klevansky}
\address{Institut f\"ur Theoretische Physik, Universit\"at Heidelberg, \\
         Philosophenweg 19, D--69120 Heidelberg, Germany}

\date{September 1995}

\maketitle
\vspace*{2cm}
\begin{abstract}
The technique of decomposing Feynman diagrams at the one loop level
into elementary integrals is generalized to the imaginary time
Matsubara formalism. The three lowest integrals, containing one, two and
three fermion lines, are provided in a form that separates out the
real and imaginary parts of these complex functions, according to the
input arguments, in a fashion that is suitable for numerical
evaluation. The forms given can be evaluated for arbitrary values of
temperature, particle mass, particle momenta and chemical potential.
\end{abstract}

\pacs{MSC numbers: 65K05, 81Q30, 81T18, 82B10}

\section{Introduction}
The imaginary time Matsubara formalism for Green functions \cite{Matsu}
is a venerated method of dealing with problems of finite temperature.
This is so because of the fact that Wick's theorem can easily be shown
to hold for the Matsubara operators. As such, this method of dealing
with finite temperature systems has found wide application, primarily
in condensed matter physics \cite{FetWa}, but also recently in problems
relevant for high energy physics. In particular, it has been recently
used in studies of effective QCD Lagrangians (see, e.~g. the reviews
\cite{SPK} and \cite{HatKun} and references cited therein), through which
an understanding of hot and dense matter is
sought\cite{HKZV,ZHK,elaste,su2hadron,su3hadron}. However, to date most
such studies have often been constrained by specific parameter choices
to enable tractability -- often degenerate particle masses, zero
chemical potential or special kinematics are chosen. Furthermore, often
only the principal values of these integrals are considered, with the
complex nature being completely ignored. An extension of those
calculations to physically relevant cases with arbitrary parameters and
generalized kinematics, as may be appropriate for example, in
calculating cross sections in an $SU(2)$ or $SU(3)$ effective model of
QCD \cite{elaste,su2hadron,su3hadron}, becomes technically extremely
complicated. It is thus the purpose of this paper to address the
technical problems associated with such integrals, and present them in
a form suitable for the numerical evaluation of both their real and
imaginary components.  This work has come about through a study of
hadronization processes within the Nambu--Jona--Lasinio model
\cite{su3hadron}, and is intended to provide the technical knowledge
required for such functions for researchers who are active in this
field. The associated computer code conformal to the analysis presented
here is freely available\footnote{Programs are available by anonymous
ftp to {\tt Trick.MPI-HD.MPG.DE}, the location of the archive file is
{\tt /pub/loopies-1.0.shar.Z}.}.

For calculations at zero temperature and density, it has been shown
\cite{OlVer}, that all occurring transition amplitudes at the one loop
level can be decomposed into a number of elementary integrals, which in
turn can be classified by the number of particle lines they contain.
Since this decomposition technique can be generalized to the Matsubara
formalism, the aim of this paper is to provide suitable forms for the
elementary integrals containing one, two and three fermion lines, which
we term $A$, $B_0$ and $C_0$, respectively, which can be used for
numerical evaluation. (The calculation can be easily taken over to
boson loops on replacing the Fermi distribution by the Bose
distribution.) We analyze these integrals, explicitly displaying their
complex nature that is a function of the input parameters. The
parameters that may then be specified can be of the most general form,
i.~e.  arbitrary temperatures, particle masses, chemical potentials and
kinematics. As such, the associated computer code can be treated as a
library--like `black box'.

We note at this point that the calculation of the one loop integral $A$
is straightforward. Its evaluation here serves merely to remind the
reader of the usage of the Matsubara formalism and it also renders this
work complete. The evaluation of $B_0$, on the other hand, serves to
demonstrate the complexities involved in a consistent evaluation of
this function. Finally we deal with the three line integral $C_0$,
which then is highly nontrivial and complicated.

It is shown in this paper that each of integrals $B_0$ and $C_0$ (and
thus implicitly all higher one loop integrals) can be written as a sum
over terms that are constructed from a single generic function that is
taken at different values of its arguments.  The analytical structure
of these generic functions is connected with the number of nonvanishing
external momenta. We are thus required to distinguish between several
cases in evaluating the generic functions when constructing a physical
quantity.

As is usually the case in relativistic quantum mechanics, the one and
two fermion line integrals $A$ and $B_0$ are divergent, which means
that they have to be regularized. We do this using a three momentum
cutoff $\Lambda$, which means that we integrate over all momenta with
$|\vec p| < \Lambda$.  The three fermion line integral $C_0$ is however
convergent.  We nevertheless perform our analysis of this function
including a three momentum cutoff. The reasons for doing this are
twofold:  (i) the extension to the case $\Lambda\to\infty$ is simple to
obtain and provides a check of our calculations, and (ii) for models
such as that of Nambu and Jona--Lasinio extended to QCD, it is regarded
as internally consistent within the model to restrict all quark momenta
in all loops.

Having dealt with the analytical complexities of the loop integrals, we
demonstrate in the last section the usage of the integral decomposition
in a simple physical example, viz. that of calculating the kaon masses
in the $SU(3)$ Nambu--Jona--Lasinio model. This involves the integrals
$A$ and $B_0$. Use of $C_0$ may be found, for example, in
Ref.~\cite{su3hadron}.

This paper is organized as follows: in Sec.~\ref{defsec}, we define the
functions $A$, $B_0$ and $C_0$. These functions are brought into a form
suitable for numerical integration in Secs.~\ref{Asec}, \ref{B0sec} and
\ref{C0sec}, respectively. At the end of Sec.~\ref{B0sec}, we also give
some technical hints for the evaluation of our results.  In
Sec.~\ref{applic}, we give an example for the application of our
calculations. Further examples can also be found in Ref.~\cite{su3hadron}.
In Sec.~\ref{sumsec} we summarize and conclude.  Some special formulae
are proven in an appendix.

\section{Definition of $A$, $B_0$ and $C_0$}
\label{defsec}
We define the functions $A$, $B_0$ and $C_0$ to be the analytical
continuation of the one, two and three fermion line integrals,
\be
 A(m, \mu, \fspace \beta, \Lambda) =
\frac{16\pi^2}{\beta} \sum_n \exp(i\omega_n\eta) \int \frac{d^3p}{(2\pi)^3}
\frac{1}{(i\omega_n+\mu)^2-E^2} \Komma
\label{Adef}
\ee
\bea
& &B_0(k, \fspace m_1, \mu_1, \fspace  m_2, \mu_2, i\nu_m,
\fspace \beta, \Lambda) =
\label{B0def} \\ & & \qquad
\frac{16\pi^2}{\beta} \sum_n \exp(i\omega_n\eta) \int \frac{d^3p}{(2\pi)^3}
\frac{1}{((i\omega_n+\mu_1)^2-E_1^2)}
\frac{1}{((i\omega_n-i\nu_m+\mu_2)^2-E_2^2)}
\nonumber \Komma
\eea and \bea
& &C_0(k, q, \delta_{\vec k,\vec q}, \fspace m_1, \mu_1, \fspace
m_2, \mu_2, i\nu_m, \fspace m_3, \mu_3,  i\alpha_l, \fspace \beta, \Lambda) =
\label{C0def} \\ & & \qquad
\frac{16\pi^2}{\beta} \sum_n \exp(i\omega_n\eta) \int \frac{d^3p}{(2\pi)^3}
\frac{1}{((i\omega_n+\mu_1)^2-E_1^2)}
\frac{1}{((i\omega_n-i\nu_m+\mu_2)^2-E_2^2)}
\nonumber \\ & & \hspace{4cm} \times
\frac{1}{((i\omega_n-i\alpha_l+\mu_3)^2-E_3^2)}
\nonumber \Komma
\eea
in which the complex frequencies $i\nu_m$ and $i\alpha_l$ are to be
continued to their values on the real axis after the Matsubara
summation on $n$ is carried out. It is to be understood that the limit
$\eta\to 0$ is to be taken after the Matsubara summation. In the above
expressions, we have introduced the abbreviations
\begin{equation}\label{espec} \begin{array}{lll}
E   = \sqrt{p^2+m^2} & \hspace{2cm} &
E_1 = \sqrt{p^2+m_1^2} \\[0.5cm]
E_2 = \sqrt{(\vec p - \vec k)^2+m_2^2} & \hspace{2cm} &
E_3 = \sqrt{(\vec p - \vec q)^2+m_3^2} \Komma
\end{array} \end{equation}
where the symbols $m_i$ and $\mu_i$ denote the masses and chemical
potentials of the fermions running on the individual lines of the
diagram. The $m_i$ are to be understood to include a negative
infinitesimal imaginary part, $m_i\to m_i-i\epsilon$, $\epsilon>0$.  In
Eqs.~(\ref{Adef}) to (\ref{C0def}), $\beta$ is the inverse temperature,
$\beta=1/T$.  The three momentum cutoff is denoted by $\Lambda$, and
all integrals are considered for $|\vec p| \le \Lambda$.

The complex four momenta $(i\nu_m; \vec k)$ and $(i\alpha_l; \vec q)$,
are the four momenta that enter the loop. The frequencies $i\nu_m$ and
$i\alpha_l$ are bosonic in nature,
\be
\nu_m = \frac{2m\pi}{\beta} \hspace{2cm} \alpha_l = \frac{2l\pi}{\beta}
\Komma
\ee
while the frequencies $i\omega_n$ are fermionic,
\be
\omega_n = \frac{(2n+1)\pi}{\beta}
\Punkt
\ee
After analytical continuation, $i\nu_m$ and $i\alpha_l$ become the
zeroth components associated with the three momenta $\vec k$ and $\vec
q$, i.~e. $i\nu_m\to k_0$, $i\alpha_l\to q_0$.  One may note at this
point that, due to rotational invariance, the functions $B_0$ and $C_0$
do not fully depend on both $\vec k$ and $\vec q$; $B_0$ depends only
on $k = |\vec k|$, while $C_0$ depends only on $k$, $q=|\vec q|$ and
the enclosed angle $\delta_{\vec k,\vec q}$, as is implied by the
arguments specified on the left hand side of Eqs.~(\ref{B0def}) and
(\ref{C0def}).

\section{The Calculation of $A$}
\label{Asec}
The calculation of $A$ is simple and straightforward.  After evaluating
the Matsubara sum by contour integration in the usual
fashion\cite{FetWa}, one has
\bea
A(m, \mu, \fspace \beta, \Lambda) &=&
     16 \pi^2 \pint \left[f(E-\mu) \frac{1}{2E} -
     f(-E-\mu)\frac{1}{2E} \right] \nonumber \\
&=& 4 \eint dE \, \sqrt{E^2-m^2} \left( f(E-\mu)-f(-E-\mu) \right)
\label{Aend} \Komma
\eea
where we have introduced the Fermi distribution function
\be f(x) = \frac{1}{\exp(\beta x) + 1} \ee
and the energy cutoff
\be \Lambda_E = \sqrt{\Lambda^2+m^2} \Punkt \ee
{}From Eq.~(\ref{Aend}), it follows that $A(m, \mu, \fspace \beta, \Lambda)
= A(m, -\mu, \fspace \beta, \Lambda)$.  This integral can be easily
performed by numerical integration. At $T=0$, it has a closed analytical
form,
\be
A(m, \mu, \fspace \infty, \Lambda) = 2\Theta(\Lambda_E^2-\mu^2)
\left(m^2 \log \frac{\Lambda+\Lambda_E}{\kappa + \kappa'}
+ \kappa\kappa'-\Lambda\Lambda_E\right) \Komma
\ee
where $\kappa=\max(m,|\mu|)$ and $\kappa'=\sqrt{\kappa^2-m^2}$.

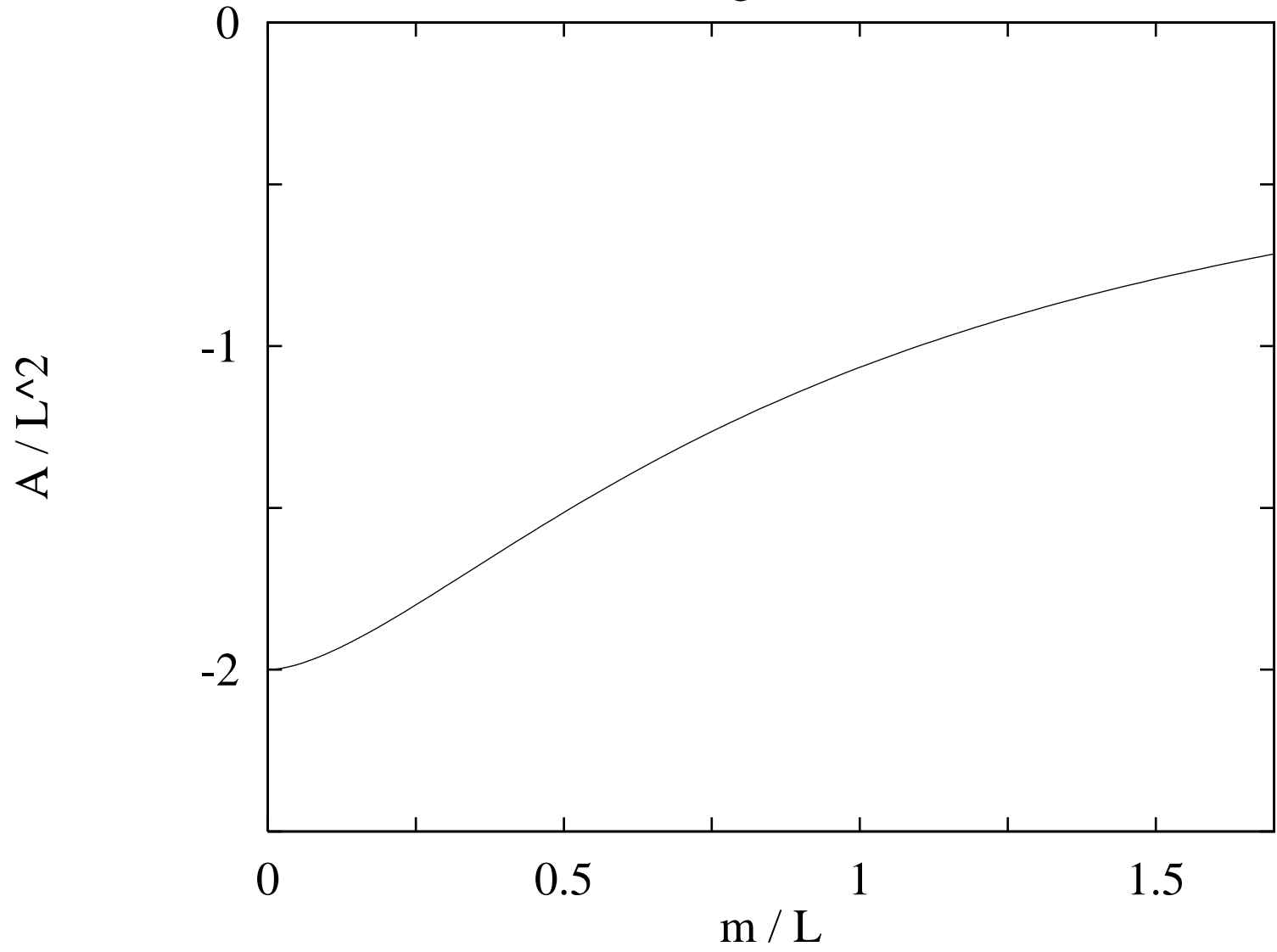
\begin{figure}
\caption{$A/\Lambda^2$ as a function of $m/\Lambda$ at $\mu=0$,
$\beta\to\infty$ and $\Lambda=602.3\MeV$.} \label{paplot}
\end{figure}

In Fig.~\ref{paplot}, we show $A/\Lambda^2$ as a function of
$m/\Lambda$, for the specific parameter choice $\mu=0$,
$\beta\to\infty$ and $\Lambda=602.3\MeV$. At $m=0$, we obtain the limit
$A(0, 0, \fspace \infty, \Lambda) = -2\Lambda^2$. For larger values of
$m$, $A$ increases continously to zero.

\section{The Calculation of $B_0$}
\label{B0sec}
\subsection{Matsubara Summation}
The two line integral $B_0(k, \fspace m_1, \mu_1, \fspace  m_2, \mu_2,
i\nu_m, \fspace \beta, \Lambda)$ can be analyzed in a similar fashion
to $A(m,\mu,\fspace\beta,\Lambda)$.  The denominator of
Eq.~(\ref{B0def}) has the four zeros
\begin{equation}
i\omega_n = - \mu_1 \pm E_1 \hspace{2cm}
i\omega_n = i\nu_m - \mu_2 \pm E_2 \Komma
\end{equation}
from which we obtain, after the summation over $n$, the terms
\bea
& & B_0(k, \fspace m_1, \mu_1, \fspace  m_2, \mu_2, i\nu_m,
\fspace \beta, \Lambda) \nonumber \\
& & \qquad = 16 \pi^2 \pint \left[
\frac{f(E_1-\mu_1)}{2E_1} \frac{1}{(-\mu_1+E_1-i\nu_m+\mu_2)^2-E_2^2}
\right. \nonumber \\
& & \qquad \phantom{= 16 \pi^2 \pint}
- \frac{f(-E_1-\mu_1)}{2E_1} \frac{1}{(-\mu_1-E_1-i\nu_m+\mu_2)^2-E_2^2}
\nonumber \\
& & \qquad \phantom{= 16 \pi^2 \pint}
+ \frac{f(E_2-\mu_2)}{2E_2} \frac{1}{(i\nu_m-\mu_2+E_2+\mu_1)^2-E_1^2}
\nonumber \\
& & \qquad \phantom{= 16 \pi^2 \pint}
- \left.
\frac{f(-E_2-\mu_2)}{2E_2} \frac{1}{(i\nu_m-\mu_2-E_2+\mu_1)^2-E_1^2}
\right] \Punkt \label{B0sum}
\eea
To eliminate the angular dependence in the arguments of the Fermi
functions, we carry out the substitution
\be \vec p \to \vec k - \vec p \label{subst} \ee
in the third and fourth terms of Eq.~(\ref{B0sum}) and obtain
\bea
& & B_0(k, \fspace m_1, \mu_1, \fspace  m_2, \mu_2, i\nu_m,
\fspace \beta, \Lambda) \nonumber \\
& & \qquad = 8\pi^2 \pint \left[
\frac{f(E_1-\mu_1)}{E_1}\frac{1}{\lambda^2 - 2\lambda E_1 + 2\vec{p}\vec{k}
-k^2 + m_1^2 - m_2^2} \right. \nonumber \\
& & \qquad \phantom{= 8\pi^2 \pint}
-\frac{f(-E_1-\mu_1)}{E_1}\frac{1}{\lambda^2 + 2\lambda E_1 + 2\vec{p}\vec{k}
-k^2 + m_1^2 - m_2^2} \nonumber \\
& & \qquad \phantom{= 8\pi^2 \pint}
+\frac{f(E_2-\mu_2)}{E_2}\frac{1}{\lambda^2 + 2\lambda E_2 + 2\vec{p}\vec{k}
-k^2 - m_1^2 + m_2^2} \nonumber \\
& & \qquad \phantom{= 8\pi^2 \pint} - \left.
\frac{f(-E_2-\mu_2)}{E_2}\frac{1}{\lambda^2 - 2\lambda E_2 + 2\vec{p}\vec{k}
-k^2 - m_1^2 + m_2^2} \right] \label{vierdrei}
\eea
where the meaning of $E_1$ and $E_2$ has changed to become
\be E_1 = \sqrt{p^2+m_1^2} \qquad \qquad E_2 = \sqrt{p^2+m_2^2} \Punkt \ee
We have also introduced
\be \lambda = i\nu_m + \mu_1 - \mu_2 \Punkt \label{lamdef} \ee
Note that $\lambda$ will be {\em real\/} after the analytical
continuation. One should further note that the substitution
Eq.~(\ref{subst}) has been performed under the assumption that $\Lambda\gg
k$. Thus no corresponding shift of variables in the integration limits has
been carried out at this point.

Without loss of generality, we may choose the coordinate system in such
a way that $\vec k$ points in the $z$ direction. Then the
integration over the angle $\phi$ yields a factor $2\pi$. It is now
useful to consider each term occurring in Eq.~(\ref{vierdrei})
separately.  These correspond to the individual poles in the original
expression and have a common generic structure. One may therefore
decompose $B_0$ into the four terms
\bea
B_0(k, \fspace m_1, \mu_1, \fspace  m_2, \mu_2, i\nu_m,
\fspace \beta, \Lambda)
&=& \tilde B_0^+(-\lambda, k, \fspace m_1, m_2, \mu_1, \fspace \beta, \Lambda)
\nonumber \\ 
&-& \tilde B_0^-(\lambda, k, \fspace m_1, m_2, \mu_1, \fspace \beta, \Lambda)
\nonumber \\ 
&+& \tilde B_0^+(\lambda, k, \fspace m_2, m_1, \mu_2, \fspace \beta, \Lambda)
\nonumber \\ 
&-& \tilde B_0^-(-\lambda, k, \fspace m_2, m_1, \mu_2, \fspace \beta, \Lambda)
\label{bzerl}
\eea
in terms of the generic function
\bea
& &\tilde B_0^\pm(\lambda, k, \fspace m, m', \mu, \fspace \beta, \Lambda) =
\nonumber \\
& &\hspace{1.5cm} 2\eint dE \, p f(\pm E-\mu)
\int_{-1}^{+1} dx \frac{1}{\lambda^2+2\lambda E + 2pkx -k^2 +m^2-m'^2}
\eea
with $p = \sqrt{E^2-m^2}$. Numerical computation of $B_0$ thus relies
purely on the evaluation of $\tilde B_0^\pm$. The
analytical structure of this integral differs for $k=0$ and $k>0$, so
that these two cases must be considered separately. We examine these in
the following two subsections.  We may also note that the symmetry
relation
\be
B_0(k, \fspace m_1, \mu_1, \fspace  m_2, \mu_2, -k_0,
\fspace \beta, \Lambda) =
B_0(k, \fspace m_2, \mu_2, \fspace  m_1, \mu_1, k_0,
\fspace \beta, \Lambda) \Komma \label{B0sym}
\ee
can be inferred from Eq.~(\ref{vierdrei}).

\subsection{Calculation of $\tilde B_0^\pm$ for $k=0$}
For $k=0$, the integral over $x$ is trivial, and we obtain
\be
\tilde B_0^\pm = 4 \eint dE
\frac{\sqrt{E^2-m^2} f(\pm E - \mu)}{\lambda^2+2\lambda E +m^2 - m'^2} \Punkt
\ee
(Here and in the following, we drop the arguments of $\tilde B_0^\pm$
for convenience.)  The analytical continuation of the Matsubara
frequency $i\nu_m$ to real values leads to poles in the integrand at
$E=E_0$, where
\be
E_0=-\frac{\lambda^2+m^2-m'^2}{2\lambda} \label{E0def} \Komma
\ee
if $m \le E_0 \le \Lambda_E$. To calculate the integral in this case,
we recall that the particle masses are complex, $m^2 \to m^2
-i\epsilon$, $m'^2 \to m'^2 -i\epsilon$, and apply the formula
\be
\lim_{\epsilon\to 0} \frac{1}{x-i\epsilon} =
{\cal P}\frac{1}{x} + i\pi\delta(x) \Komma \label{celeb}
\ee
where ${\cal P}$ denotes the Cauchy principal value. From this, we find
\bea
\tilde B_0^\pm &=& \lim_{\epsilon \to 0} 4 \eint dE
\frac{p f(\pm E - \mu)}{\lambda^2 + 2\lambda E + m^2 - m'^2 - i\epsilon \sgn
(\lambda)} \nonumber \\
&=& \lim_{\epsilon \to 0} \frac{2}{\lambda} \eint dE
\frac{p f(\pm E - \mu)}{E + \frac{\lambda^2+m^2-m'^2}{2\lambda} - i\epsilon}
\nonumber \\
&=& 4 \Vp_m^{\Lambda_E} dE
\frac{p f(\pm E - \mu)}{\lambda^2+2\lambda E + m^2 -m'^2} \nonumber \\
&+& i\frac{2\pi}{\lambda} p_0 f(\pm E_0 -\mu)
\Theta((\Lambda_E-E_0)(E_0-m))
\label{B00end} \Komma
\eea
where $p_0=\sqrt{E_0^2-m^2}$ and the $\Theta$ function ensures that the
imaginary part appears only if $E_0$ lies within the integration interval
$[m,\Lambda_E]$.

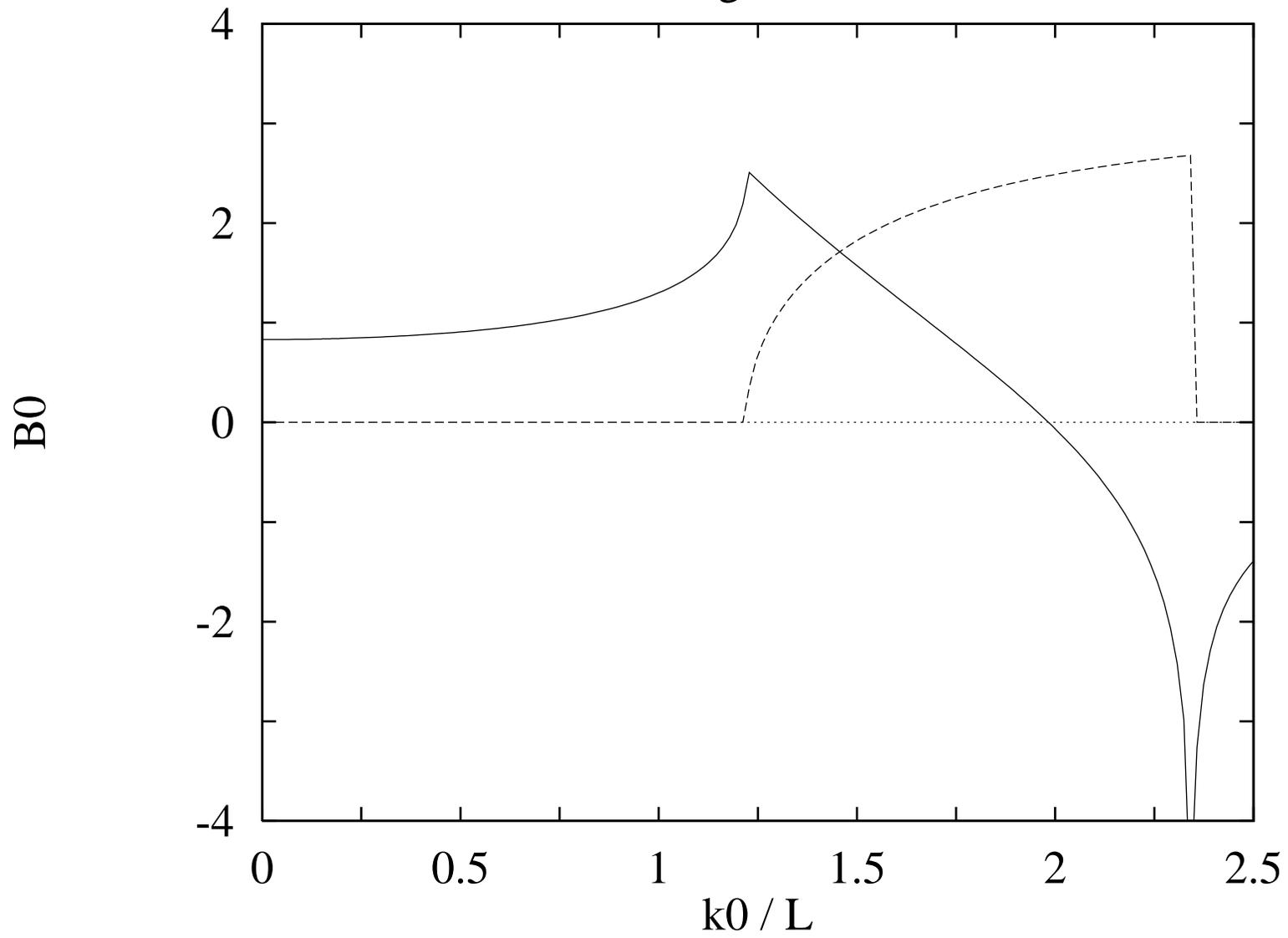
\begin{figure}
\caption[]{$B_0$ for $k=0$, $\mu_1=\mu_2=0$, $m_1=m_2=367.7\MeV$,
$T=0$ and $\Lambda=602.3\MeV$. The solid line gives the real part, the
dashed line the imaginary part.} \label{pb0plot}
\end{figure}

Figure~\ref{pb0plot} shows $B_0$ as a function of $k_0/\Lambda$ for the
parameter set $k=0$, $\mu_1=\mu_2=0$, $m_1=m_2=m=367.7\MeV$,
$\beta\to\infty$, $\Lambda=602.3\MeV$. One recognizes that at small
$k_0$, $B_0$ is a real function, whereas for $k_0\ge 2m$ an imaginary
part emerges. At $k_0>2\Lambda_E$, the imaginary part vanishes due to
the presence of the cutoff.

\subsection{Calculation of $\tilde B_0^\pm$ for $k>0$}
For $k>0$, the generic function $\tilde B_0^\pm$ has the form
\bea
& &\tilde B_0^\pm(\lambda, k, \fspace m, m', \mu, \fspace \beta, \Lambda) =
2\limit \eint dE\, p f(\pm E-\mu) \nonumber \\ & & \hspace{2cm} \times
\int_{-1}^{+1} dx \frac{1}{\lambda^2+2\lambda E + 2pkx -k^2 +m^2-m'^2
-i\epsilon\sgn(\lambda)} \Komma
\eea
where we have already indicated that the infinitesimal imaginary part
in the integrand is determined by the sign of $\lambda$.  In the limit
$\epsilon\to 0$, the integrand contains a singularity at
\be
x = - \frac{1}{2pk}(\lambda^2+2\lambda E -k^2 +m^2-m'^2)
\ee
if
\be
|\lambda^2+2\lambda E -k^2 +m^2-m'^2| < 2pk \Punkt \label{singsing}
\ee
On performing the $x$ integration and applying Eq.~(\ref{celeb}), one
obtains
\bea
\tilde B_0^\pm &=& \lim_{\epsilon \to 0}
\frac{1}{k}\eint dE\, f(\pm E-\mu)
\int_{-1}^{+1} dx \frac{1}{x + \frac{\lambda^2+2\lambda E -k^2 +m^2-m'^2}{2pk}
-i\epsilon\sgn(\lambda)}
\label{B0knoend} \\
&=& \frac{1}{k} \eint dE\, f(\pm E-\mu) \log \left|
\frac{(\lambda+E)^2-(p-k)^2-m'^2}{(\lambda+E)^2-(p+k)^2-m'^2}\right|
\nonumber \\ &+&
i\frac{\pi\sgn(\lambda)}{k} \eint dE\,
\Theta(2pk - |\lambda^2+2\lambda E -k^2 +m^2-m'^2|) f(\pm E-\mu)
\Punkt \nonumber
\eea
The integral for the imaginary part can be done analytically:
\bea
\int dE \frac{1}{\exp(\pm\beta E -\beta\mu)+1}
= E \mp \frac{1}{\beta}\log(1+\exp(\pm\beta E -\beta\mu)) \Punkt
\eea

The integrand for the real part in Eq.~(\ref{B0knoend}) contains
logarithmic poles at the energies $E=E_{1/2}$ which are defined by
\be
|\lambda^2+2\lambda E_{1/2} -k^2 +m^2-m'^2| = 2k\sqrt{E_{1/2}^2-m^2}
\label{e12def} \Punkt
\ee
These energies also constitute the limits of integration for the
imaginary part.  Squaring Eq.~(\ref{e12def}) leads to a quadratic
equation for $E_{1/2}$, which has the solutions
\be
E_{1/2} = \frac{\lambda(\lambda^2-k^2+m^2-m'^2)}{2(k^2-\lambda^2)}
\pm k \frac{\sqrt{(\lambda^2-k^2+m^2-m'^2)^2-4m^2(k^2-\lambda^2)}}
{2(k^2-\lambda^2)} \label{schnitt} \Punkt
\ee
There are three possible cases for Eq.~(\ref{schnitt}):
\begin{enumerate}
\item No solutions exist for $E_{1/2}$, or both solutions lie outside
      the integration interval $[m,\Lambda_E]$. Since there is at least
      one point ($E=m$), where the $\Theta$ function in
      Eq.~(\ref{B0knoend}) vanishes, we conclude that in this case the
      imaginary part disappears.
\item One solution $E_1$ lies inside the integration interval, the other
      one, $E_2$ lies outside. It follows that the integral for the
      imaginary part has to be performed from $E_1$ to $\Lambda_E$ in
      this case.
\item The two solutions $E_1 < E_2$ lie within the integration interval. In
      this case, the integral for the imaginary part has to be performed
      from $E_1$ to $E_2$.
\end{enumerate}
Since we will deal with integration limits of this type very often, we
will adopt the convention that the solutions $E_{1/2}$ of
Eq.~(\ref{schnitt}) are tailored as described above.
Eq.~(\ref{B0knoend}) then takes the form
\bea
\tilde B_0^\pm
&=& \frac{1}{k} \eint dE\, f(\pm E-\mu) \log \left|
\frac{(\lambda+E)^2-(p-k)^2-m'^2}{(\lambda+E)^2-(p+k)^2-m'^2}\right|
\label{B0kend} \\ &+&
i\frac{\pi\sgn(\lambda)}{k} \int_{E_1}^{E_2} dE\, f(\pm E-\mu)
\Punkt \nonumber
\eea

Together with Eq.~(\ref{schnitt}), Eq.~(\ref{B0kend}) provides a
sufficient basis for the numerical calculation of $\tilde B_0^\pm$. It
can be shown that in the limit $k\to 0$, Eq.~(\ref{B0kend}) continously
approaches Eq.~(\ref{B00end}). Thus the generic function $\tilde
B_0^\pm$ is determined, and consequently $B_0$ via Eq.~(\ref{bzerl}).

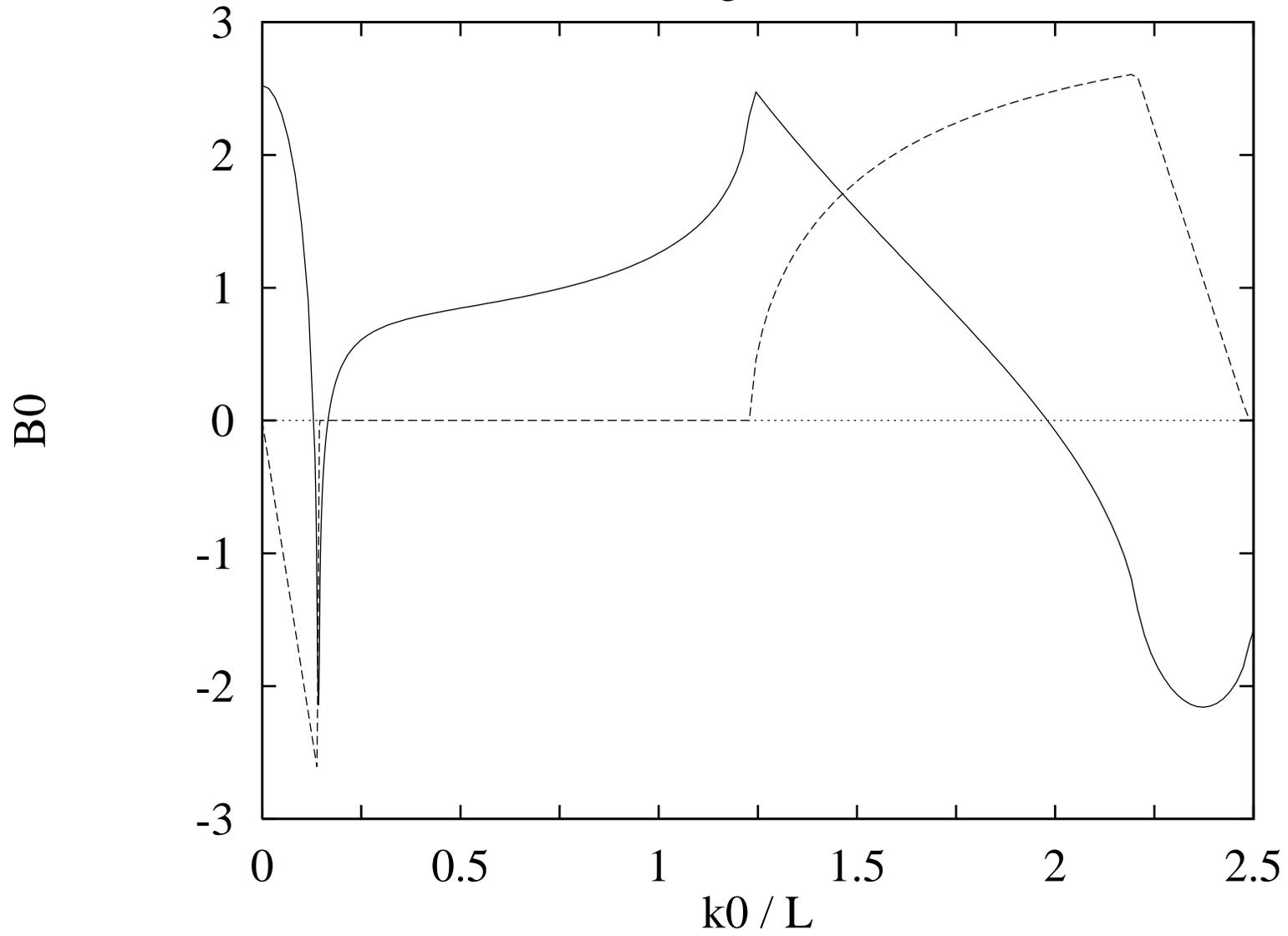
\begin{figure}
\caption[]{$B_0$ for $k=100\MeV$, $\mu_1=\mu_2=0$, $m_1=m_2=367.7\MeV$,
$T=0$ and $\Lambda=602.3\MeV$. The solid line gives the real part, the
dashed line the imaginary part.} \label{pbkplot}
\end{figure}

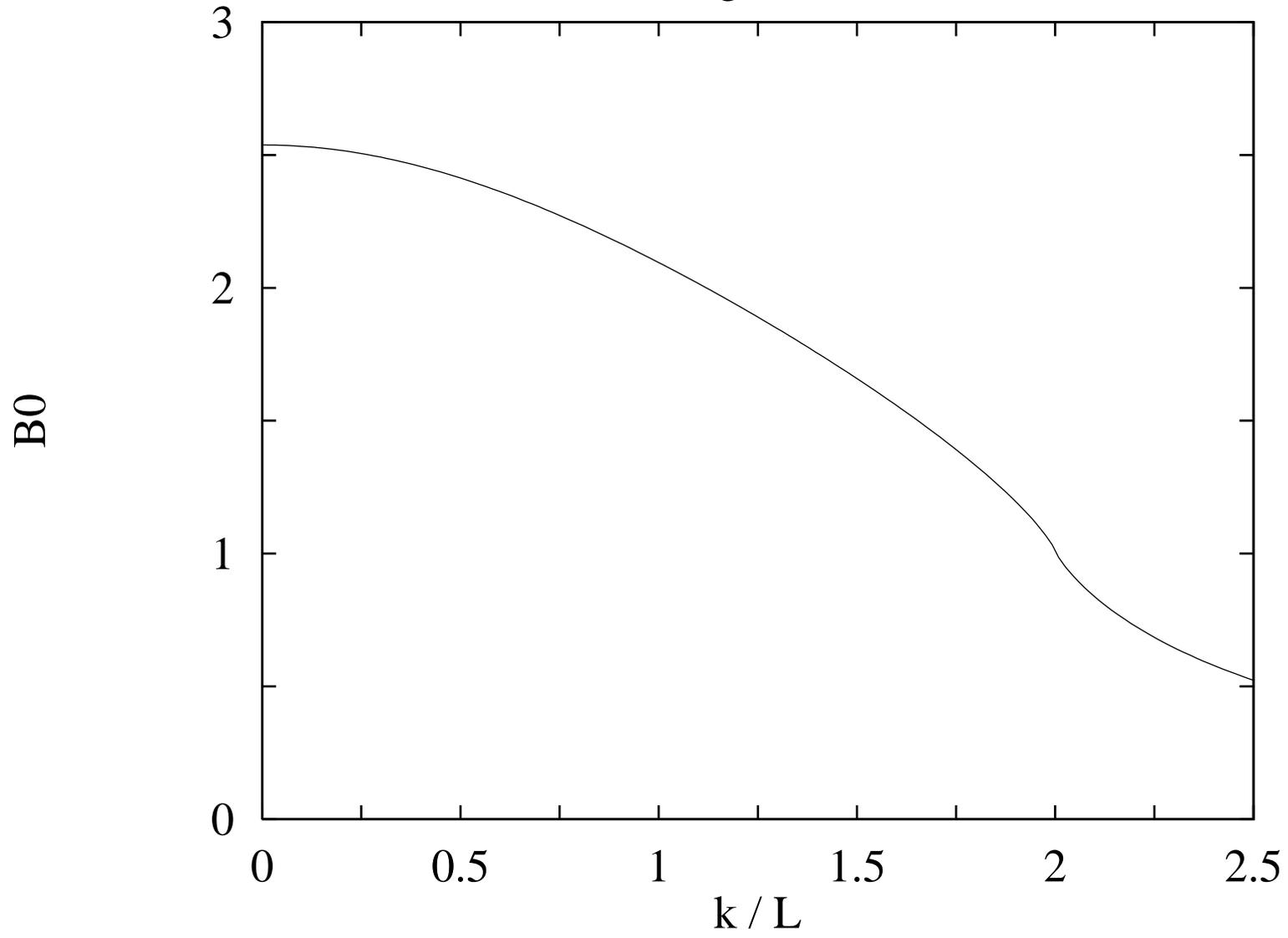
\begin{figure}
\caption[]{$B_0$ for $k_0=0$, $\mu_1=\mu_2=0$, $m_1=m_2=367.7\MeV$,
$T=0$ and $\Lambda=602.3\MeV$.  Note that $B_0$ is a real function in
this case.} \label{pbsplot}
\end{figure}

Figure~\ref{pbkplot} shows $B_0$ for the same parameter set as was used
in Fig.~\ref{pb0plot}, but with $k=100\MeV$. For $k_0 < k$,
Eq.~(\ref{schnitt}) has one solution, from which one obtains an
imaginary part. For $k_0>2m$, Eq.~(\ref{schnitt}) has two solutions,
which become cut off at high $k_0$, so that the imaginary part
continously goes to zero. Figure~\ref{pbsplot} shows another special case.
Here we plot $B_0$ as a function of $k$ for $k_0=0$. In this case, we
have no imaginary part at all\cite{elaste}.

\subsection{Programming Considerations} \label{procon}
At this point, we deviate from the analytical calculation to discuss
some of the basic aspects of programming such integrals. Comments made
here are relevant for the $B_0$ and $C_0$ integrals, the first of which
is the least complicated integral containing multiple singularities. We
discuss this now, bearing in mind that similar techniques are required
for the more complex $C_0$ integral.

To calculate $B_0$ for the most general case via the decomposition
(\ref{bzerl}), one has to take several things into consideration.
First, the type of singularity determines the type of integration
routine that should be utilized.  One requires three such routines: (i)
one routine which integrates smooth functions, (ii) one which
integrates functions containing logarithmic or other integrable
singularities, and (iii) one which computes the Cauchy principal value
of an integral containing a $1/x$ singularity. Examples for this can be
found in the literature \cite{NumRec,Quadpack}.

On entry, the program has to decide, which of the forms (\ref{B00end}),
pertinent to the case $k=0$, or (\ref{B0kend}), pertinent for $k>0$, to
use. It is necessary to handle the two cases separately, since (i)
Eq.~(\ref{B00end}) and Eq.~(\ref{B0kend}) have a different analytical
structure, and (ii) Eq.~(\ref{B0kend}) contains a factor $1/k$.  In
actual practice, the case $k=0$ is needed very often.  After that, the
position of singularities can be computed according to Eq.~(\ref{E0def})
or Eq.~(\ref{schnitt}), respectively. The appropriate integration
routine has to be chosen, depending on the result of this computation.

Note that although the decomposition as given Eq.~(\ref{bzerl}) is the
most general one possible, it does not always provide the optimal way
of calculating $B_0$. If, for example, one knows a priori that an actual
calculation will be confined to $m_1=m_2$, $\mu_1=\mu_2$ and $k=0$, the
four terms of Eq.~(\ref{bzerl}) can be combined, yielding the
simplified expression
\bea
& &B_0(0, \fspace m, \mu, \fspace  m, \mu, k_0, \fspace \beta, \Lambda)
= 8 \Vp_m^{\Lambda_E} dE (f(E-\mu)-f(-E-\mu)) \frac{p}{k_0^2-4E^2} \\
& & \hspace{1.5cm} - i \pi
\left(f\left(\frac{k_0}{2}-\mu\right)-f\left(-\frac{k_0}{2}-\mu\right)\right)
\frac{\sqrt{k_0^2-4m^2}}{k_0} \Theta((2\Lambda_E-k_0)(k_0-2m)) \Punkt
\nonumber
\eea
The result of this equation stays finite in the limit $k_0\to 0$. The
functions $\tilde B_0^\pm$, however, individually {\it diverge} in this
limit, so that a computation of $B_0$ by decomposition suffers from
cancellations at small $k_0$. Note that similar simplifications follow
for $C_0$.

\section{The Calculation of $C_0$}
\label{C0sec}
\subsection{Matsubara Summation}
In this section, we examine the analytical structure of the three
fermion line integral $C_0$, and decompose it into a sum involving a
single generic function.

The three fermion line function given in Eq.~(\ref{C0def}) has poles at
\begin{mathletters} \bea
i\omega_n &=& -\mu_1 \pm E_1 \\
i\omega_n &=& i\nu_m -\mu_2 \pm E_2 \\
i\omega_n &=& i\alpha_l -\mu_3 \pm E_3
\eea\end{mathletters}
with $E_1$, $E_2$ and $E_3$ as specified in Eq.~(\ref{espec}).
In analogy to Eq.~(\ref{lamdef}), we define the complex variables
\begin{mathletters} \bea
\lambda_1 &=& i \nu_m + \mu_1 - \mu_2 \\
\lambda_2 &=& i \alpha_l + \mu_1 - \mu_3 \\
\lambda_3 &=& i \nu_m - i \alpha_l - \mu_2 + \mu_3 = \lambda_1 - \lambda_2
\Punkt
\eea \end{mathletters}
We again note that these parameters become real numbers after the
analytical continuation is performed.

After summing the Matsubara frequencies, one obtains a sum over six
terms that arise from each of the poles,
\bea
& &C_0(k, q, \delta_{\vec k, \vec q}, \fspace m_1, \mu_1,
\fspace m_2, \mu_2, i\nu_m,
\fspace m_3, \mu_3,  i\alpha_l, \fspace \beta, \Lambda) \nonumber \\
& & \qquad = 8 \pi^2 \pint \left[
\frac{f(E_1-\mu_1)}{E_1}\frac{1}{[(\lambda_1-E_1)^2-E_2^2][(\lambda_2-E_1)^2
-E_3^2]}
\right. \nonumber \\
& &\qquad \phantom{= 8 \pi^2 \pint} + (E_1 \to -E_1) \nonumber \\
& &\qquad \phantom{= 8 \pi^2 \pint} +
\frac{f(E_2-\mu_2)}{E_2}\frac{1}{[(\lambda_1+E_2)^2-E_1^2][(\lambda_3+E_2)^2
-E_3^2]} \nonumber \\
& &\qquad \phantom{= 8 \pi^2 \pint} + (E_2 \to -E_2) \nonumber \\
& &\qquad \phantom{= 8 \pi^2 \pint} +
\frac{f(E_3-\mu_3)}{E_3}\frac{1}{[(\lambda_2+E_3)^2-E_1^2][(\lambda_3-E_3)^2
-E_2^2]} \nonumber \\
& &\left. \qquad \phantom{= 8 \pi^2 \pint} + (E_3 \to -E_3)
\right] \Punkt
\eea
It is useful to make the substitution $\vec p \to \vec k - \vec p$ in
the third and fourth terms, $\vec p \to \vec q - \vec p$ in the fifth
and sixth terms, in order to get rid of the angular dependence in the
arguments of the Fermi functions.  Again this is performed under the
assumption that $\Lambda\gg k,q$.  After this, we obtain a
decomposition of $C_0$ of the form
\bea
& &C_0(k, q, \delta_{\vec k, \vec q},
\fspace m_1, \mu_1, \fspace m_2, \mu_2, i\nu_m, \fspace m_3, \mu_3,
i\alpha_l, \fspace \beta, \Lambda) \nonumber \\ & & \qquad \qquad
= \tilde C_0^+(-\lambda_1, -\lambda_2, \fspace k, q, \delta_{\vec k, \vec q},
\fspace m_1, m_2, m_3, \mu_1, \fspace \beta, \Lambda)
\nonumber \\ & & \qquad \qquad
- \tilde C_0^-(\lambda_1, \lambda_2, \fspace k, q, \delta_{\vec k, \vec q},
 \fspace m_1, m_2, m_3, \mu_1, \fspace \beta, \Lambda)
\nonumber \\ & & \qquad \qquad
+ \tilde C_0^+(\lambda_1, \lambda_3, \fspace k, |\vec{k}-\vec{q}|,
\delta_{\vec k, \vec{k}-\vec{q}},
\fspace m_2, m_1, m_3, \mu_2, \fspace \beta, \Lambda)
\nonumber \\ & & \qquad \qquad
- \tilde C_0^-(-\lambda_1, -\lambda_3, \fspace k, |\vec{k}-\vec{q}|,
\delta_{\vec k, \vec{k}-\vec{q}},
\fspace m_2, m_1, m_3, \mu_2, \fspace \beta, \Lambda)
\nonumber \\ & & \qquad \qquad
+ \tilde C_0^+(\lambda_2, -\lambda_3, \fspace q, |\vec{q}-\vec{k}|,
\delta_{\vec q, \vec{q}-\vec{k}},
\fspace m_3, m_1, m_2, \mu_3, \fspace \beta, \Lambda)
\nonumber \\ & & \qquad \qquad
- \tilde C_0^-(-\lambda_2, \lambda_3, \fspace q, |\vec{q}-\vec{k}|,
\delta_{\vec q, \vec{q}-\vec{k}},
\fspace m_3, m_1, m_2, \mu_3, \fspace \beta, \Lambda) \Komma
\label{C0decomp}
\eea
where we have introduced the three fermion line generic function
\bea & &
\tilde C_0^\pm (\lambda_1, \lambda_2, \fspace k, q, \delta,
\fspace m, m_1, m_2, \mu, \fspace \beta, \Lambda) = 8\pi^2
\label{C0ansatz} \\ \times
& & \hspace{0.5cm} \pint \frac{f(\pm E - \mu)}{E}
\frac{1}{[(\lambda_1+E)^2-(\vec{p}-\vec{k})^2-m_1^2]
         [(\lambda_2+E)^2-(\vec{p}-\vec{q})^2-m_2^2]} \nonumber
\eea
with $E=\sqrt{p^2+m^2}$. In analogy to Eq.~(\ref{B0sym}), one can
derive a symmetry relation for the function $C_0$, as being
\bea
& &C_0(k, q, \delta_{\vec k, \vec q},
\fspace m_1, \mu_1, \fspace m_2, \mu_2, i\nu_m, \fspace m_3, \mu_3,
i\alpha_l, \fspace \beta, \Lambda) \\ & & \hspace{1.5cm} =
C_0(k, |\vec k - \vec q|, \delta_{\vec k, \vec k - \vec q},
\fspace m_2, \mu_2, \fspace m_1, \mu_1, -i\nu_m, \fspace m_3, \mu_3,
i\alpha_l-i\nu_m, \fspace \beta, \Lambda) \nonumber \Punkt
\eea

As was the case for the two line generic function $\tilde B_0^\pm$,
$\tilde C_0^\pm$ is calculated for several special cases. We list these
in order of increasing difficulty:
\begin{enumerate}
\item The case $k=q=0$.
\item The case $k=0$, $q>0$.
\item The case $k>0$, $q>0$, $\vec p$ and $\vec k$ collinear.
\item The case $k>0$, $q>0$, $\vec p$ and $\vec k$ not collinear
      -- general case.
\end{enumerate}
In the following sections, we will derive results for each of these
special cases.

\subsection{Calculation of $\tilde C_0^\pm$ for $k=q=0$}
In this case, Eq.~(\ref{C0ansatz}) reduces to
\bea
\tilde C_0^\pm &=& 8\pi^2 \pint \frac{f(\pm E - \mu)}{E}
\frac{1}{[(\lambda_1+E)^2-p^2-m_1^2] [(\lambda_2+E)^2-p^2-m_2^2]}
\nonumber \\
&=& \frac{1}{\lambda_1\lambda_2} \limit \eint dE
\frac{pf(\pm E - \mu)}{(E-E_1-i\epsilon)(E-E_2-i\epsilon)} \Komma
\label{c0001}
\eea
where we have again omitted the arguments of $\tilde C_0^\pm$ for
convenience and have included a small imaginary part in the
denominators to do the analytical continuation.  The constants $E_1$
and $E_2$ are defined by
\begin{mathletters} \bea
E_1 &=& - \frac{\lambda_1^2+m^2-m_1^2}{2\lambda_1} \label{E1def0} \\
E_2 &=& - \frac{\lambda_2^2+m^2-m_2^2}{2\lambda_2} \label{E2def0} \Punkt
\eea\end{mathletters}
If $m \le E_1=E_2 \le \Lambda_E$, the integral in
Eq.~(\ref{c0001}) diverges. We suppose therefore that
$E_1\ne E_2$, if one of them lies inside the integration interval.
After taking the limit $\epsilon\to 0$, we obtain
\bea
\tilde C_0^\pm &=& 4 \Vp_m^{\Lambda_E} dE \frac{pf(\pm E-\mu)}
{[(\lambda_1+E)^2-p^2-m_1^2] [(\lambda_2+E)^2-p^2-m_2^2]} \label{c000end} \\
&+& i\frac{\pi}{\lambda_1\lambda_2(E_1-E_2)}
\Big(p_1 f(\pm E_1 - \mu) \Theta((\Lambda_E - E_1)(E_1-m))
\nonumber \\ & & \hspace{4cm}
- p_2 f(\pm E_2 - \mu) \Theta((\Lambda_E - E_2)(E_2-m))\Big) \Punkt
\nonumber
\eea
The $\Theta$ functions here guarantee that the imaginary part occurs
only if the singularities appear inside the integration interval. As
was detailed in Sec.~\ref{procon}, the numerical evaluation of the
integral in Eq.~(\ref{c000end}) has to proceed differently if $E_1$ or
$E_2$ lie within the interval $[m,\Lambda_E]$ (Cauchy integration), or
if they lie outside of $[m,\Lambda_E]$ (integration of a smooth
function).

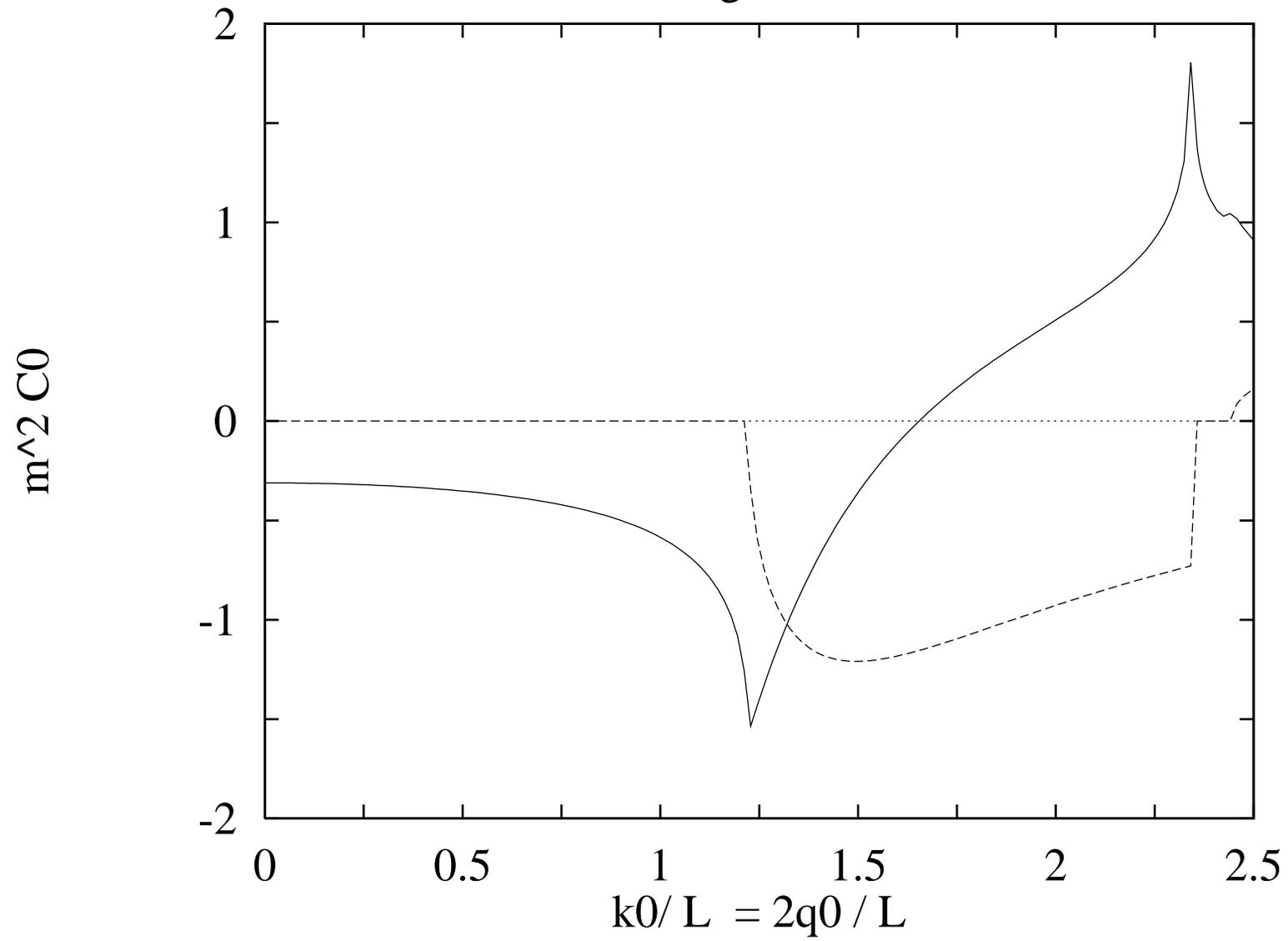
\begin{figure}
\caption[]{$m^2C_0$ for $k=q=0$, $m_1=m_2=m_3=367.7\MeV$,
$\mu_1=\mu_2=\mu_3=0$, $q_0=2k_0$, $T=0$ and $\Lambda=602.3\MeV$.  The
solid line gives the real part, the dashed line the imaginary part.}
\label{pc0plot}
\end{figure}

Figure~\ref{pc0plot} shows $m^2C_0$ for $k=q=0$ as a function of
$k_0/\Lambda=2q_0/\Lambda$ for a similar parameter set as was used in
Fig.~\ref{pb0plot}.  Again we obtain an imaginary part for $k_0>2m$.

\subsection{Calculation of $\tilde C_0^\pm$ for $k=0$, $q>0$}
Without loss of generality, we assume that $\vec q$ points in the
$z$ direction. With this assumption, the scalar product $\vec p\vec q$
occurring in Eq.~(\ref{C0ansatz}) simplifies to $pq\cos\theta$. The
integration over $\phi$ becomes trivial and one obtains
\bea
\tilde C_0^\pm &=& \frac{1}{\pi} \limit \int d^3p \frac{f(\pm E-\mu)}{E}
\frac{1}{\lambda_1^2+2\lambda_1E+m^2-m_1^2-i\epsilon\sgn(\lambda_1)}
\nonumber \\ & & \qquad \qquad \times
\frac{1}
{\lambda_2^2+2\lambda_2E+2pq\cos\theta-q^2+m^2-m_2^2-i\epsilon\sgn(\lambda_2)}
\nonumber \\
&=&\frac{1}{2\lambda_1\lambda_2} \limit \eint dE
\frac{pf(\pm E-\mu)}{E-E_1-i\epsilon}\int_{-1}^{+1}dx \frac{1}
{E-E_2+\frac{pq}{\lambda_2}x-i\epsilon} \label{c001int} \Komma
\eea
where now
\be
E_2 = - \frac{\lambda_2^2-q^2+m^2-m_2^2}{2\lambda_2} \label{E2defk}
\ee
and $E_1$ is as defined in Eq.~(\ref{E1def0}).
The angular integral is singular if $|\lambda_2(E-E_2)|<pq$.
This equation has exactly the same structure as Eq.~(\ref{singsing}),
so Eq.~(\ref{schnitt}) and the following discussion can be applied
here too. We label the endpoints of the singular interval $E_{21}$ and
$E_{22}$. The angular integral becomes
\bea
& &\limit\int_{-1}^{+1}dx \frac{1}{E-E_2+\frac{pq}{\lambda_2}x-i\epsilon}
=\frac{\lambda_2}{pq} \limit \int_{-1}^{+1}dx\frac{1}
{\frac{\lambda_2(E-E_2)}{pq}+x-i\epsilon\sgn(\lambda_2)}
\nonumber \\ & & \hspace{3cm}
= \frac{\lambda_2}{pq} \log \left|
\frac{\frac{\lambda_2(E-E_2)}{pq}+1}
{\frac{\lambda_2(E-E_2)}{pq}-1}\right|
+i\frac{\pi\lambda_2\sgn(\lambda_2)}{pq}
\Theta\left(1-\left|\frac{\lambda_2(E-E_2)}{pq}\right|\right)
\nonumber \\ & & \hspace{3cm}
= \frac{\lambda_2}{pq} \log \left|
\frac{(\lambda_2+E)^2-(p-q)^2-m_2^2}{(\lambda_2+E)^2-(p+q)^2-m_2^2}
\right| \nonumber \\ & & \hspace{5cm}
+ i\frac{\pi\lambda_2\sgn(\lambda_2)}{pq}
\Theta((E_{22}-E)(E-E_{21})) \Punkt
\eea
For $\tilde C_0^\pm$ we have accordingly
\bea
\tilde C_0^\pm
&=& \frac{1}{2\lambda_1q} \limit \eint dE \frac{f(\pm E-\mu)}{E-E_1-i\epsilon}
\log \left|\frac{(\lambda_2+E)^2-(p-q)^2-m_2^2}{(\lambda_2+E)^2-(p+q)^2-m_2^2}
\right|
\nonumber \\
&+& i\frac{\pi\sgn(\lambda_2)}{2\lambda_1q} \limit
\int_{E_{21}}^{E_{22}} dE \frac{f(\pm E-\mu)}{E-E_1-i\epsilon} \Punkt
\eea
In this expression, we again have to take the limit $\epsilon\to 0$.
The final expression we find is
\bea
\tilde C_0^\pm &=& \frac{1}{2\lambda_1q} \Vp_m^{\Lambda_E} dE
\frac{f(\pm E-\mu)}{E-E_1}\log \left| \frac{(\lambda_2+E)^2-(p-q)^2-m_2^2}
{(\lambda_2+E)^2-(p+q)^2-m_2^2} \right|
\nonumber \\ &-& \frac{\pi^2\sgn(\lambda_2)}{2\lambda_1q}f(\pm E_1-\mu)
\Theta((E_{22}-E_1)(E_1-E_{21}))
\nonumber \\ &+& i\frac{\pi f(\pm E_1-\mu)}{2\lambda_1q} \log \left|
\frac{(\lambda_2+E_1)^2-(p_1-q)^2-m_2^2}{(\lambda_2+E_1)^2-(p_1+q)^2-m_2^2}
\right| \Theta((\Lambda_E-E_1)(E_1-m))
\nonumber \\ &+& i\frac{\pi\sgn(\lambda_2)}{2\lambda_1q}
\Vp_{E_{21}}^{E_{22}} dE \frac{f(\pm E-\mu)}{E-E_1}
\label{c001end} \Punkt
\eea
A numerical evaluation of Eq.~(\ref{c001end}) has to take into account
Cauchy singularities at $E=E_1$ and (integrable) logarithmic
singularities at $E=E_{21}$ and $E=E_{22}$.

\subsection{Calculation of $\tilde C_0^\pm$ for $k>0$, $q>0$,
            $\vec q$ and $\vec k$ collinear}
Again, we assume without loss of generality that $\vec q$ and
$\vec k$ have the form
\be
\vec q = \left( \begin{array}{c} 0 \\ 0 \\ q \end{array} \right)
\qquad \qquad
\vec k = \left( \begin{array}{c} 0 \\ 0 \\ \eta k \end{array} \right)
\ee
where $\eta=\pm 1$. With this assumption, we have $\vec p \vec q =
pq\cos\theta$ and $\vec p \vec k = \eta pk\cos\theta$ in
Eq.~(\ref{C0ansatz}).  The $\phi$ integration is again trivial, and one
obtains
\bea
\tilde C_0^\pm &=& \frac{1}{\pi} \limit \int d^3p \frac{f(\pm E-\mu)}{E}
\frac{1}{\lambda_1^2+2\lambda_1E+2\eta pk\cos\theta-k^2+m^2-m_1^2
-i\epsilon\sgn(\lambda_1)}
\nonumber \\ & & \qquad \times
\frac{1}{\lambda_2^2+2\lambda_2E+2pq\cos\theta-q^2+m^2-m_2^2
-i\epsilon\sgn(\lambda_2)}
\nonumber \\ &=& \frac{1}{2\lambda_1\lambda_2} \limit
\eint dE\, pf(\pm E-\mu)
\nonumber \\ & & \hspace{3cm} \times
\int_{-1}^{+1} dx \frac{1}{E-E_1+\frac{\eta pk}{\lambda_1}x-i\epsilon}
\frac{1}{E-E_2+\frac{pq}{\lambda_2}x-i\epsilon} \label{gl98}
\eea
with
\be
E_1 = - \frac{\lambda_1^2-k^2+m^2-m_1^2}{2\lambda_1}
\ee
and $E_2$ from Eq.~(\ref{E2defk}).
The integral over $x$ becomes (cf. \cite{GroeHof1}, integral No. 12.8)
\bea
& &\limit \int_{-1}^{+1} dx
\frac{1}{E-E_1+\frac{\eta pk}{\lambda_1}x-i\epsilon}
\frac{1}{E-E_2+\frac{pq}{\lambda_2}x-i\epsilon}
\nonumber \\ & & \qquad =
\frac{\lambda_1\lambda_2}{pq\lambda_1(E-E_1)-\eta pk\lambda_2(E-E_2)}
\nonumber \\ & & \qquad \qquad \qquad \times
\log \left| \frac{(\lambda_2(E-E_2)+pq)(\lambda_1(E-E_1)-\eta pk)}
{(\lambda_2(E-E_2)-pq)(\lambda_1(E-E_1)+\eta pk)} \right|
\nonumber \\ & & \qquad \qquad
+ i \frac{\pi\lambda_1\lambda_2\eta\sgn(\lambda_1)
\Theta\left(1-\left|\frac{\lambda_1(E-E_1)}{pk}\right|\right)}
{\eta pk\lambda_2(E-E_2)-pq\lambda_1(E-E_1)}
\nonumber \\ & & \qquad \qquad
+ i \frac{\pi\lambda_1\lambda_2\sgn(\lambda_2)
\Theta\left(1-\left|\frac{\lambda_2(E-E_2)}{pq}\right|\right)}
{pq\lambda_1(E-E_1)-\eta pk\lambda_2(E-E_2)} \Punkt \label{winkelin}
\eea
In what follows, we label the energies for which
\be |\lambda_1(E-E_1)| = pk \ee
is fulfilled $E_{1\,1/2}$,
the energies for which
\be |\lambda_2(E-E_2)| = pq \Punkt \ee
is fulfilled $E_{2\,1/2}$.  These can be calculated explicitly via
Eq.~(\ref{schnitt}). Note that the discussion following
Eq.~(\ref{schnitt}) applies here too.

The second and third terms of Eq.~(\ref{winkelin}) still contain a
singularity at $E=E_0$, where
\be E_0=\frac{2}{\zeta}(q\lambda_1 E_1-\eta k\lambda_2 E_2) \ee
with \be \zeta = 2 (q\lambda_1-\eta k\lambda_2) \Punkt \ee
This is, however, not true for the first term, since the logarithmic
factor goes to zero at $E=E_0$, making the integrand continous here. So
we again add a small imaginary part in the denominators where
appropriate and obtain
\bea
\tilde C_0^\pm &=& \frac{1}{\zeta} \eint dE \frac{f(\pm E-\mu)}{E-E_0}
\log \left| \frac{(\lambda_2(E-E_2)+pq)(\lambda_1(E-E_1)-\eta pk)}
{(\lambda_2(E-E_2)-pq)(\lambda_1(E-E_1)+\eta pk)} \right|
\nonumber \\ &-& i \frac{\pi\eta\sgn(\lambda_1)}{\zeta}
\limit \int_{E_{11}}^{E_{12}} dE \frac{f(\pm E-\mu)}
{E-E_0-i\epsilon}
+ i \frac{\pi\sgn(\lambda_2)}{\zeta}
\limit \int_{E_{21}}^{E_{22}} dE \frac{f(\pm E-\mu)}
{E-E_0-i\epsilon}
\nonumber \\
&=& \frac{1}{\zeta} \eint dE \frac{f(\pm E-\mu)}{E-E_0}
\log \left| \frac{(\lambda_2(E-E_2)+pq)(\lambda_1(E-E_1)-\eta pk)}
{(\lambda_2(E-E_2)-pq)(\lambda_1(E-E_1)+\eta pk)} \right|
\nonumber \\ &+& \frac{\pi^2\eta\sgn(\lambda_1)}{\zeta}
f(\pm E_0-\mu) \Theta((E_{12}-E_0)(E_0-E_{11}))
\nonumber \\ &-& \frac{\pi^2\sgn(\lambda_2)}{\zeta}
f(\pm E_0-\mu) \Theta((E_{22}-E_0)(E_0-E_{21}))
\nonumber \\ &-& i \frac{\pi\eta\sgn(\lambda_1)}{\zeta}
\Vp_{E_{11}}^{E_{12}} dE \frac{f(\pm E-\mu)}{E-E_0}
+ i \frac{\pi\sgn(\lambda_2)}{\zeta}
\Vp_{E_{21}}^{E_{22}} dE \frac{f(\pm E-\mu)}{E-E_0}
\label{c0koll} \Punkt
\eea
For the numerical evaluation of the real part, one has to take into
account the logarithmic poles at $E=E_{1\,1/2}$ and $E=E_{2\,1/2}$. For
the imaginary part, the Cauchy singularity at $E=E_0$ must be
integrated.

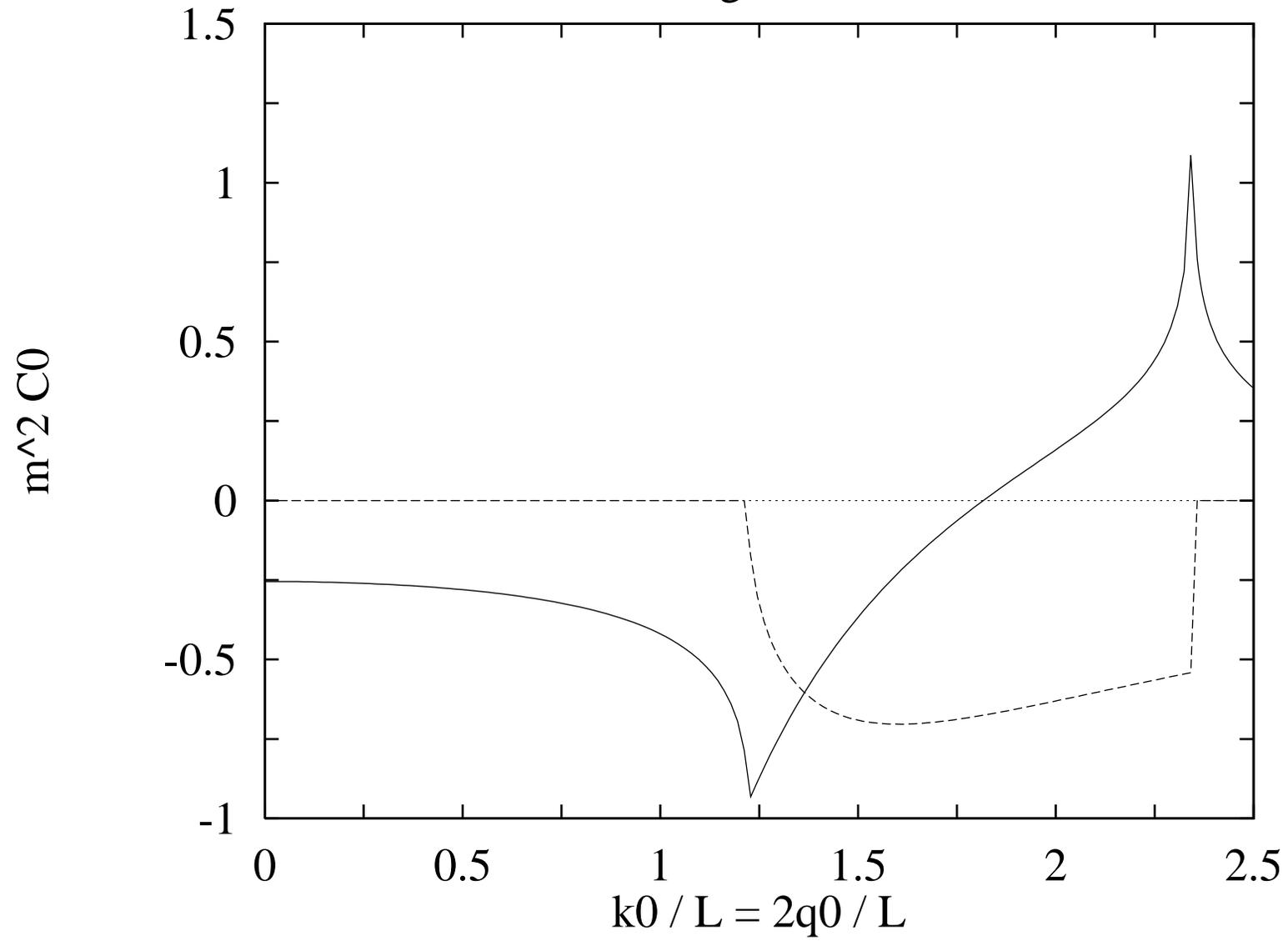
\begin{figure}
\caption[]{$m^2C_0$ for $k=0$, $q=q_0=2k_0$, $m_1=m_2=m_3=367.7\MeV$,
$\mu_1=\mu_2=\mu_3=0$, $T=0$ and $\Lambda=602.3\MeV$.  The solid line
gives the real part, the dashed line the imaginary part.}
\label{pc2plot}
\end{figure}

In Fig.~\ref{pc2plot}, we show $m^2C_0$ as a function of $k_0/\Lambda$
for $q=q_0=k_0/2$ and all other parameters as in Fig.~\ref{pc0plot}.
This special case has been considered previously in
Ref.~\cite{su2hadron}.  As can be seen from Eq.~(\ref{C0decomp}), the
computation of $C_0$ requires a computation of $\tilde C_0^\pm$ for
$k=0$, $q>0$, as well as for $\vec k$, $\vec q$ collinear.

\subsection{Calculation of $\tilde C_0^\pm$ for the General Case}
In this subsection, we handle the general case of arbitrary momenta
in the evaluation of $\tilde C_0^\pm$ from Eq.~(\ref{C0ansatz}).
We choose the coordinate system in such a way that
\be
\vec q = \left( \begin{array}{c} 0 \\ 0 \\ q \end{array} \right)
\qquad \qquad
\vec k = \left( \begin{array}{c} 0 \\ k\sin\delta \\ k\cos\delta \end{array}
\right) \Punkt
\ee
{}From this choice we obtain
\begin{mathletters} \bea
\vec{p}\vec{k} &=& pk(\cos\theta\cos\delta+\sin\theta\sin\delta\cos\phi) \\
\vec{p}\vec{q} &=& pq\cos\theta \Punkt
\eea \end{mathletters}
It is useful to introduce the abbreviations
\begin{mathletters} \label{c12def} \bea
c_1 &=& \frac{\lambda_1^2+2\lambda_1E+m^2-m_1^2-k^2}{2pk} \\
c_2 &=& \frac{\lambda_2^2+2\lambda_2E+m^2-m_2^2-q^2}{2pq} \Punkt
\eea \end{mathletters}

In terms of these variables, $\tilde C_0^\pm$ takes the form
\bea
\tilde C_0^\pm &=& \frac{1}{4\pi kq} \limit \eint dE \frac{f(\pm E - \mu)}{p}
\int_0^\pi d\theta \frac{\sin\theta}{c_2+\cos\theta-i\epsilon\sgn(\lambda_2)}
\nonumber \\ & & \qquad \times
\int_0^{2\pi} d\phi
\frac{1}{c_1+\cos\theta\cos\delta+\sin\theta\sin\delta\cos\phi
-i\epsilon\sgn(\lambda_1)} \Punkt
\eea
The additional difficulty introduced by the generality of handling
arbitrary momenta is evident in the $\phi$ integration. As will be
seen in the following, this can however still be dealt with
analytically.

To perform the $\phi$ integration, we use the formula
\be
\lim_{\epsilon\to 0}
\int_0^{2\pi}\frac{d\phi}{a+b\cos\phi-i\epsilon} =
\frac{2\pi}{\sqrt{|a^2-b^2|}}
\left(\Theta(a^2-b^2)\sgn(a)+i\Theta(b^2-a^2)\right) \Komma \label{PHIINT}
\ee
which we prove in Appendix~\ref{appphi}. With this formula, one
immediately obtains the result
\be
\tilde C_0^\pm =\frac{1}{2kq}\eint dE \frac{f(\pm E - \mu)}{p}
\int_{-1}^{+1}dx \frac{\Theta(\Delta_1)\sgn(c_1+x\cos\delta)+i\Theta(-\Delta_1)
\sgn(\lambda_1)}{\sqrt{|\Delta_1|}(x+c_2-i\epsilon\sgn(\lambda_2))}
\label{c0phi}
\ee
where
\be
\Delta_1 = (c_1+x\cos\delta)^2-\sin^2\delta(1-x^2) \Punkt
\ee
In the following subsections, we examine the remaining angular integrals
that are required.

\subsubsection{The $\theta$ Integration for $\Delta_1<0$}
In the next step, we calculate the integral
\be
\int_{-1}^{+1}dx \frac{\Theta(-\Delta_1) \sgn(\lambda_1)}
{\sqrt{|\Delta_1|}(x+c_2-i\epsilon\sgn(\lambda_2))} \Punkt
\ee
We first note that for $x=\pm 1$, $\Delta_1$ is always nonnegative.
$\Delta_1$ has zeros at
\begin{mathletters} \label{dzero} \be
x = x_1 = -c_1\cos\delta - \sqrt{1-c_1^2}\sin\delta
\ee and \be
x = x_2 = -c_1\cos\delta + \sqrt{1-c_1^2}\sin\delta
\ee\end{mathletters}
with $-1\le x_1\le x_2\le +1$.
Consequently two conditions must be fulfilled in order to obtain
$\Delta_1\le 0$: (i) $|c_1|<1$ and (ii) $x_1\le x\le x_2$. From this
it follows that we can apply the formula
\be
\lim_{\epsilon\to 0}
\int_a^b\frac{dx}{\sqrt{(b-x)(x-a)}(x+c-i\epsilon)} =
\frac{\pi}{\sqrt{|\Delta|}}\left(\overline{\sgn(x+c)}
\Theta(\Delta)+i\Theta(-\Delta)\right) \label{FORMEL} \Komma
\ee
where $\Delta=(a+c)(b+c)$, which is proven in Appendix~\ref{appformel}.
Note that for $\Delta>0$ $\sgn(x+c)$ is a constant in the interval
$[a,b]$, which is expressed by the symbol $\overline{\sgn(x+c)}$.

Using Eq.~(\ref{FORMEL}), one obtains
\bea
& &\int_{-1}^{+1} \frac{\Theta(-\Delta_1) \sgn(\lambda_1)}
{\sqrt{|\Delta_1|}(x+c_2-i\epsilon\sgn(\lambda_2))}dx
\\ & & \qquad
= \Theta(1-c_1^2)
\frac{\pi\sgn(\lambda_1)}{\sqrt{|\Delta_0|}}\left(\sgn(c_2-c_1\cos\delta)
\Theta(\Delta_0)+i\sgn(\lambda_2)\Theta(-\Delta_0)\right) \Komma \nonumber
\eea
where the factor $\overline{\sgn(x+c_2)}$ has been taken at
$x = (x_1+x_2)/2=-c_1\cos\delta$. $\Delta_0$ can be obtained from $\Delta_1$
by substituting $-c_2$ for $x$:
\be
\Delta_0 = c_1^2 + c_2^2 - 2c_1c_2\cos\delta - \sin^2\delta
\ee

\subsubsection{The $\theta$ Integration for $\Delta_1>0$}
It remains to compute the integral
\bea
& &\limit \int_{-1}^{+1} \frac{\Theta(\Delta_1)\sgn(c_1+x\cos\delta)}
{\sqrt{\Delta_1}(x+c_2-i\epsilon\sgn(\lambda_2))}dx
= \Vp_{-1}^{+1} \frac{\Theta(\Delta_1)\sgn(c_1+x\cos\delta)}
{\sqrt{\Delta_1}(x+c_2)}dx \nonumber \\ & & \hspace{5cm} +
i \frac{\pi\sgn(\lambda_2)\sgn(c_1-c_2\cos\delta)}
{\sqrt{\Delta_0}} \Theta(\Delta_0) \Theta(1-c_2^2) \Punkt \label{DGint}
\eea
In doing so, one easily notes that $c_1+x\cos\delta$ vanishes only at
points where $\Delta_1\le 0$, so that the sgn term can be treated
effectively as a constant.  The real part can be computed from
\cite{GroeHof2}, integral No.~231.10.  Bearing the definition of
$x_{1/2}$ from Eq.~(\ref{dzero}) in mind, one obtains
\bea
& & \Vp_{-1}^{+1} \frac{\Theta(\Delta_1)\sgn(c_1+x\cos\delta)}
{\sqrt{\Delta_1}(x+c_2)}dx \nonumber \\ & & \hspace{2cm} =
\frac{\Theta(\Delta_0)}{\sqrt{\Delta_0}}
\bigg[ \sgn(c_1+\cos\delta) \bigg[ F(+1)-\Theta(1-c_1^2)
F(x_2)\bigg] \nonumber \\ & & \hspace{3.5cm} -
\sgn(c_1-\cos\delta) \bigg[ F(-1)-\Theta(1-c_1^2)
F(x_1)\bigg] \bigg] \nonumber \\ & & \hspace{2cm} +
\frac{\Theta(-\Delta_0)}{\sqrt{-\Delta_0}}
\bigg[\sgn(c_1+\cos\delta)\bigg[ G(+1)-
G(x_2)\bigg] \nonumber \\ & & \hspace{3.5cm} -
\sgn(c_1-\cos\delta)\bigg[ G(-1)- G(x_1)\bigg] \bigg]
\label{DGloes}
\eea
where
\be
G(x) = \arccos\frac{(c_2-c_1\cos\delta)(x+c_2)-\Delta_0}{\sin\delta
\sqrt{1-c_1^2}|x+c_2|}
\ee
and $F$ is either one of the functions
\be
F_\pm(x) = \pm\log\left|\frac{(x+c_2)(c_1\cos\delta-c_2)+\Delta_0
\mp\sqrt{\Delta_0\Delta_1}} {x+c_2}\right| \Punkt
\ee
Eq.~(\ref{DGloes}) can be further simplified by
\begin{mathletters} \be
F_\pm(x_1) = F_\pm(x_2) = \pm\log\left(\sin\delta\sqrt{1-c_1^2}\right)
\ee \be
G(x_1) = G(x_2) = 0 \Punkt
\ee \end{mathletters}

Up until this point, the calculation in this section has been purely
analytical. However, although $F_+$ and $F_-$ are equivalent in
specifying the indefinite integral pertinent to Eq.~(\ref{DGint}), they
are not equally well behaved numerically if either $|c_1|$ or $|c_2|$
is approximately equal to one. It is therefore necessary at this point
to devise a numerically stable procedure, and we give our algorithm
here. To this end, it is useful to define $\Xi$ as
\bea
\Xi &=& \sgn(c_1+\cos\delta) \bigg[ F(+1)-\Theta(1-c_1^2) F(x_2)\bigg]
\nonumber \\ &-&
\sgn(c_1-\cos\delta) \bigg[ F(-1)-\Theta(1-c_1^2) F(x_1)\bigg] \Punkt
\eea
Using the identity
\be
F_+(x)-F_-(x)=\log\left(\sin^2\delta\left|1-c_1^2\right|\right) \Komma
\ee
one can show that the following algorithm gives the correct result:
\begin{itemize} \begin{mathletters} \label{xicalc}
\item If $||c_2|-1| < ||c_1|-1|$: Compute $\Xi$ as
      \bea
      \Xi &=& \sgn(c_1+\cos\delta)F_-(+1)-\sgn(c_1-\cos\delta)F_-(-1)
            \nonumber \\ &+&
            \Theta(\cos^2\delta-c_1^2)\sgn(\cos\delta)
            \log\left|\sin^2\delta(1-c_1^2)\right|
      \eea
\item If $||c_2|-1| > ||c_1|-1|$:
      \begin{itemize}
      \item If $c_2-c_1\cos\delta>0$: Compute $\Xi$ as
            \bea
            \Xi &=& \sgn(c_1+\cos\delta)F_+(+1)-\sgn(c_1-\cos\delta)F_-(-1)
                  \nonumber \\ &-&
                  \Theta(c_1^2-\cos^2\delta)\sgn(c_1)
                  \log\left|\sin^2\delta(1-c_1^2)\right|
            \eea
      \item If $c_2-c_1\cos\delta<0$: Compute $\Xi$ as
            \bea
            \Xi &=& \sgn(c_1+\cos\delta)F_-(+1)-\sgn(c_1-\cos\delta)F_+(-1)
                  \nonumber \\ &+&
                  \Theta(c_1^2-\cos^2\delta)\sgn(c_1)
                  \log\left|\sin^2\delta(1-c_1^2)\right|
            \eea
      \end{itemize}
\end{mathletters} \end{itemize}
This is implemented in the numerical procedure.

\subsubsection{Putting the Parts together}
The final expression for $\tilde C_0^\pm$ is now given. We may write
\bea
\tilde C_0^\pm &=& \frac{1}{2kq} \eint dE
\frac{f(\pm E-\mu)}{p\sqrt{|\Delta_0|}} \Bigg( \Theta(\Delta_0)  \Xi
- \Theta(-\Delta_0)\pi\sgn(\lambda_1\lambda_2)
\\ & & \hspace{1cm} +
\Theta(-\Delta_0)\Big(\sgn(c_1+\cos\delta) G(+1) -
\sgn(c_1-\cos\delta) G(-1)\Big)
\nonumber \\ & & \hspace{1cm} +
i\pi\Theta(\Delta_0)\Big(\sgn(\lambda_1)\Theta(1-c_1^2)
\sgn(c_2-c_1\cos\delta) \nonumber \\ & & \hspace{3cm} +
\sgn(\lambda_2)\Theta(1-c_2^2) \sgn(c_1-c_2\cos\delta) \Big) \Bigg)
\nonumber \Komma
\eea
where $\Xi$ has to be taken from Eq.~(\ref{xicalc}). In the spirit of
Sec.~\ref{procon}, in order to evaluate this integral, one has to
identify the singular points of the integrand. It is easy to see that
the following points are singularities:
\begin{itemize}
\item Energies, at which $|c_1|=1$ or $|c_2|=1$. These can be calculated
      from Eq.~(\ref{schnitt}).
\item Energies, at which $\Delta_0=0$. To compute these energies, one
      notes that $c_1$ and $c_2$ can be written as
      \be
      c_1 = \frac{a_1 + b_1 E}{p}
      \qquad \qquad
      c_2 = \frac{a_2 + b_2 E}{p} \Komma
      \ee
      where $a_i$ and $b_i$ can be obtained from Eq.~(\ref{c12def}).
      With these constants, the equation $\Delta_0=0$ can be cast
      into the form
      \bea
      0 &=& E^2(b_1^2+b_2^2-2b_1b_2\cos\delta-\sin^2\delta) \nonumber \\
        &+& 2E(a_1b_1+a_2b_2-(a_1b_2+a_2b_1)\cos\delta) \nonumber \\
        &+& (a_1^2+a_2^2-2a_1a_2\cos\delta+m^2\sin^2\delta)
      \eea
      which can be easily solved for $E$.
\item Energies, at which $|c_1|=|\cos\delta|$. These can be computed by
      \be
      E=\frac{1}{2(k^2\cos^2\delta-\lambda_1^2)} \left(\lambda_1a_1\pm
      k\cos\delta\sqrt{a_1^2-4m^2(\lambda_1^2-k^2\cos^2\delta)}
      \right) \Punkt
      \ee
\end{itemize}
All of these points correspond to integrable singularities.
On closer inspection, one also obtains that the energies, at which
the arguments of the logarithmic function (in $F_\pm$) vanish, or the
energies, at which the argument of the arccosine (in $G$) becomes greater
than one, are included in these cases.

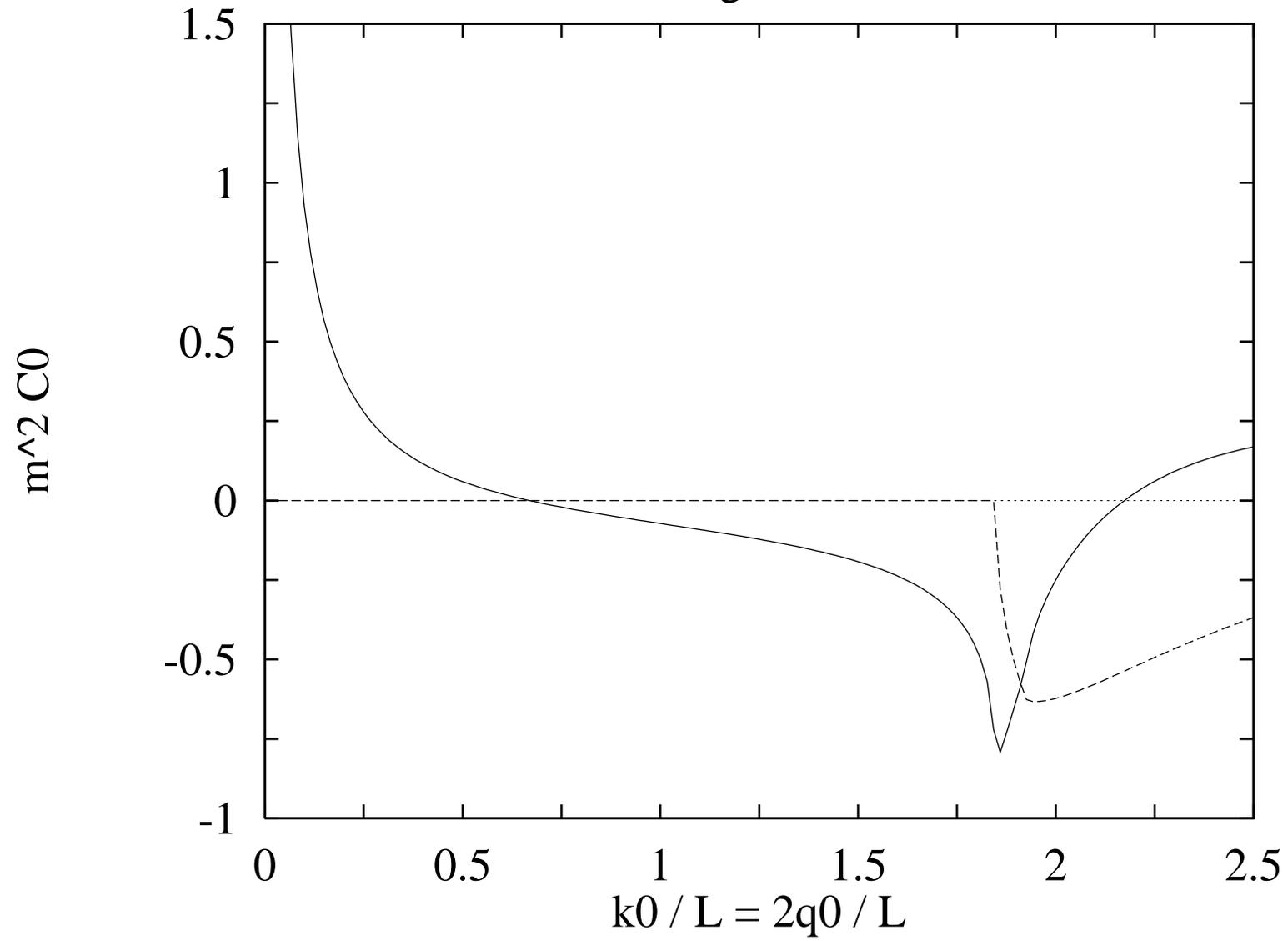
\begin{figure}
\caption[]{$m^2C_0$ for $k=3/4 k_0$, $q=3/4 q_0$, $\delta=\pi/6$,
$m_1=m_2=m_3=367.7\MeV$, $\mu_1=\mu_2=\mu_3=0$, $T=0$ and
$\Lambda=602.3\MeV$.  The solid line gives the real part, the dashed
line the imaginary part.} \label{pc4plot}
\end{figure}

To illustrate $C_0$ for the general case, we plot $m^2C_0$ in
Fig.~\ref{pc4plot} as a function of $k_0/\Lambda$ with $k=3/4k_0$,
$q=3/4q_0$ and $\delta=\pi/6$. All other parameters are as in
Fig.~\ref{pc0plot}.

\section{Application: Kaon Masses in the NJL Model}
\label{applic}
In this section, we give a specific example for a calculation
using our elementary integrals. It is also intended for the benefit
of our reader, who wishes to compare his/her results with our numbers.
To this end, we compute the kaon masses in the framework of the $SU(3)$
Nambu--Jona--Lasinio model\cite{SPK,HatKun,su3hadron} as a function of
temperature. In this model, the quark masses have to be determined from
the coupled gap equations
\begin{mathletters}\be
m_q = m_{0q} + 4GN_c i{\rm tr}_\gamma S^q(x,x) +
2KN_c^2 (i{\rm tr}_\gamma S^q(x,x)) (i{\rm tr}_\gamma S^s(x,x))
\ee \be
m_s = m_{0s} + 4GN_c i{\rm tr}_\gamma S^s(x,x) +
2KN_c^2 (i{\rm tr}_\gamma S^q(x,x)) (i{\rm tr}_\gamma S^q(x,x))
\ee \end{mathletters}
with the current quark masses $m_{0q}$, $m_{0s}$, the number of colors
$N_c$ and the coupling constants $G$ and $K$. $S^f$ denotes the finite
temperature quark propagator
\be
S^f(\vec x - \vec x', \tau - \tau') = \frac{i}{\beta}\sum_n
e^{-i\omega_n(\tau-\tau')} \int \frac{d^3p}{(2\pi)^3}
\frac{e^{i\vec p(\vec x-\vec x')}}
{\gamma_0(i\omega_n+\mu_f) - \vec\gamma \vec p -m_f}
\ee
The trace of the propagator can be expressed in terms of the function
$A$:
\be
i{\rm tr}_\gamma S^f(x,x) = - \frac{m_f}{4\pi^2}A(m_f, \mu_f)
\ee
and the gap equation takes the form
\begin{mathletters}\label{gapeq}\be
m_q = m_{0q} - \frac{N_c}{\pi^2}m_q A(m_q, \mu_q, \fspace \beta, \Lambda)
             \left( G - \frac{KN_c}{8\pi^2}
             m_s A(m_s, \mu_s, \fspace \beta, \Lambda) \right)
\ee \be
m_s = m_{0s} - \frac{GN_c}{\pi^2}m_s A(m_s, \mu_s, \fspace \beta, \Lambda)
             + \frac{KN_c^2}{8\pi^4}\left( m_q
             A(m_q, \mu_q, \fspace \beta, \Lambda)\right)^2
\ee \end{mathletters}

Meson masses are computed from the quark-antiquark scattering matrix.
The scattering matrix for light and strange quarks in the pseudoscalar
channel can be written as
\be
M = \frac{2K_{\rm eff}}{1-4K_{\rm eff}\Pi(k_0,k)} \Komma
\ee
where $(k_0,\vec k)$ is the total four momentum of the quarks,
$\Pi(k_0,k)$ the irreducible polarization and $K_{\rm eff}$ an
effective coupling constant, which is given by
\be
K_{\rm eff} = G + \frac{KN_c}{2}i{\rm tr}_\gamma S^q(x,x)
            = G - \frac{KN_c}{8\pi^2}
                m_q A(m_q, \mu_q, \fspace \beta, \Lambda)
\label{cpl45}
\ee

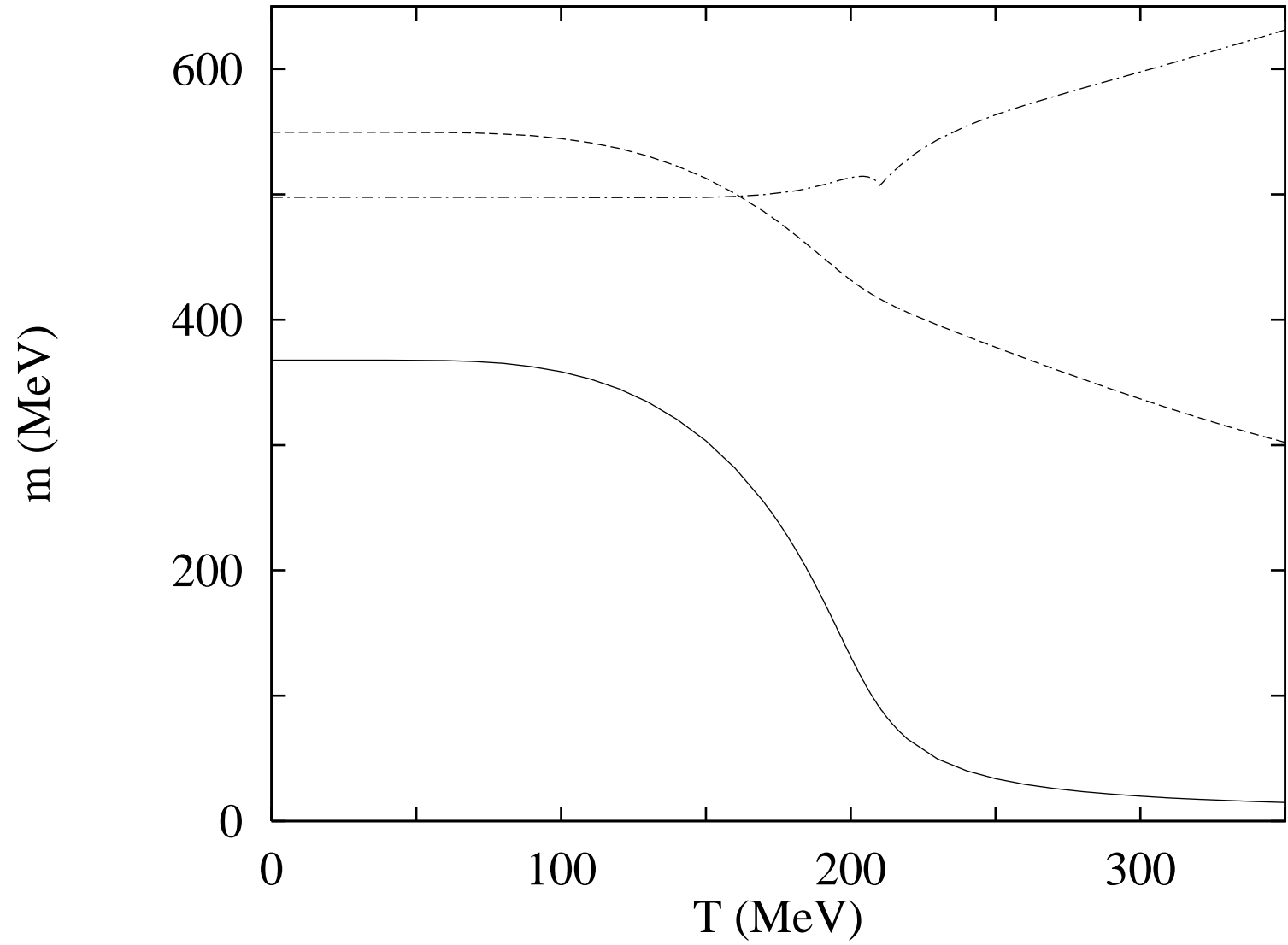
\begin{figure}
\caption[]{Temperature dependence of $m_q$ (solid line),
$m_s$ (dashed line) and $m_K$ (dot--dashed line).}
\label{massplot}
\end{figure}

The irreducible polarization $\Pi(k_0,k)$ is computed from the
diagram in Fig.~\ref{polar}. One obtains
\bea
& &-i\Pi(i\nu_m,k)
= -N_c \frac{i}{\beta}\sum_n \int \frac{d^3p}{(2\pi)^3}
{\rm tr}_\gamma \left[ iS^q(i\omega_n,\vec p)
i\gamma_5 iS^s(i\omega_n-i\nu_m,\vec p-\vec k) i\gamma_5 \right]
\nonumber \\ & & \quad
= 4iN_c \frac{1}{\beta}\sum_n \int \frac{d^3p}{(2\pi)^3}
\frac{(i\omega_n + \mu_q)(i\omega_n-i\nu_m+\mu_s) - \vec p(\vec p - \vec k)
-m_qm_s}{\left[(i\omega_n+\mu_q)^2-E_q^2\right]
\left[(i\omega_n-i\nu_m+\mu_s)^2 -E_s^2\right] }
\eea
where $E_q=\sqrt{p^2+m_q^2}$ and $E_s=\sqrt{(\vec p - \vec k)^2 + m_s^2}$.
To cast this into a form which contains the functions $A$ and $B_0$,
one uses the identity
\bea
& & (i\omega_n+\mu_q)(i\omega_n-i\nu_m+\mu_s)-
\vec p (\vec p - \vec k) - m_q m_s \nonumber \\
& & \qquad = \frac{1}{2} \bigg[\left((i\omega_n+\mu_q)^2-E_q^2\right) +
\left((i\omega_n-i\nu_m+\mu_s)^2-E_s^2\right)
\nonumber \\ & & \qquad \qquad +
\left( (m_q-m_s)^2 - (\mu_q-\mu_s+i\nu_m)^2 + k^2\right)\bigg]
\eea
After continuing $i\nu_m$ to $k_0$, one immediately obtains the result
\bea
\Pi(k_0,k) &=& -\frac{N_c}{8\pi^2}
\bigg[A(m_q, \mu_q, \fspace \beta, \Lambda)+
A(m_s, \mu_s, \fspace \beta, \Lambda) \\ &+&
\left((m_q-m_s)^2-(k_0+\mu_q-\mu_s)^2+ k ^2\right)
B_0(k, \fspace m_q, \mu_q, \fspace m_s, \mu_s, k_0,
\fspace \beta, \Lambda) \bigg] \Punkt
\nonumber
\eea
The kaon mass is now computed using the dispersion relation
\be
1-4K_{\rm eff}\Pi(m_K,0) = 0 \label{dispers} \Punkt
\ee
Using the numbers $m_{0q}=5.5\MeV$, $m_{0s}=140.7\MeV$,
$\mu_q=\mu_s=0$, $\Lambda=602.3\MeV$, $G\Lambda^2=1.835$ and
$K\Lambda^5=12.36$, one obtains the zero temperature results
$m_q=367.7\MeV$ and $m_s=549.5\MeV$ from Eq.~(\ref{gapeq}). Solving
Eq.~(\ref{dispers}) with these numbers yields $m_K=497.7\MeV$.  The
temperature dependence of $m_q$, $m_s$ and $m_K$ is computed using the
same formulae and the result is depicted in Fig.~\ref{massplot}.
Further examples can be found in Ref.~\cite{su3hadron}.

\section{Summary and Conclusion}
\label{sumsec}
In this paper, we have presented the technical aspects required for a
numerical evaluation of the one, two and three fermion line integrals
at finite temperature, structuring each as a sum of terms of a specific
generic integral taken at different values of its arguments, that are
often required for field theoretic calculations such as within the
Nambu--Jona--Lasinio model. Both real and imaginary parts of these
functions are explicitly calculated. Concomitantly, we have illustrated
all functions graphically for certain parameter sets, in order to
enable the user to verify his/her own calculation. A simple example,
employing two of the integrals has also been shown.  A computer source
code is also available that routinizes these integrals. The
generalization to bosonic line integrals can also be seen to follow in
an analogous fashion.

\section*{Acknowledgments}
We would like to thank J.~H\"ufner for his encouragement and support
of this project, and those many people who have requested that we make
this work generally available. We are also indebted to the
Bundesministerium f\"ur Bildung und Wissenschaft and the Deutsche
Forschungsgemeinschaft, who have supported this project under
contract numbers 06~HD~742 and Hu~233/4-3 respectively.

\begin{appendix}

\section{Proof of Eq.~(\ref{PHIINT})}
\label{appphi}
Our assertion is
\be
\lim_{\epsilon\to 0}
\int_0^{2\pi}\frac{d\phi}{a+b\cos\phi-i\epsilon} =
\frac{2\pi}{\sqrt{|a^2-b^2|}}
\left(\Theta(a^2-b^2)\sgn(a)+i\Theta(b^2-a^2)\right) \Punkt
\ee
The real part of this equation can be easily obtained from integral tables.
(See e.~g. \cite{GroeHof2}, integral No.~331.41.) The singularity in the
integral only appears for $|b|>|a|$, which gives the $\Theta$ function
for the imaginary part. In this case, singularities appear at
\be
\phi = \arccos(- a / b) \qquad \mbox{and} \qquad
\phi = 2\pi - \arccos(-a / b) \Punkt
\ee
The first of these singularities appears in the interval $[0,\pi]$,
the second one in the interval $[\pi,2\pi]$. We split the integral
into two parts
\bea
\lim_{\epsilon\to 0}
\int_0^{2\pi}\frac{d\phi}{a+b\cos\phi-i\epsilon} &=&
\lim_{\epsilon\to 0}
\frac{1}{b}\int_0^{\pi}\frac{d\phi}{a/b+\cos\phi-i\epsilon\sgn(b)}
\nonumber \\ &+&
\lim_{\epsilon\to 0}
\frac{1}{b}\int_\pi^{2\pi}\frac{d\phi}{a/b+\cos\phi-i\epsilon\sgn(b)}
\label{split}
\eea
and focus on the first of these two parts. Let $\phi_0= \arccos(-a /
b)$ and the function $f(\phi)$ be defined by
\be
f(\phi) = \left\{ \begin{array}{lll}
(\cos\phi + a/b)/(\phi - \phi_0) & \qquad \mbox{for} \qquad & \phi \ne
\phi_0 \\[0.5cm] \sin\phi_0 = \sqrt{1-(a/b)^2} & \qquad \mbox{for} \qquad &
\phi = \phi_0
\end{array} \right.  \Punkt
\ee
It can be easily seen that $f(\phi)$ is continous in the interval $[0,\pi]$
and
\bea
\lim_{\epsilon\to 0}
\frac{1}{b}\int_0^{\pi}\frac{d\phi}{a/b+\cos\phi-i\epsilon\sgn(b)}
&=& \lim_{\epsilon\to 0} \frac{1}{b}\int_0^{\pi}\frac{d\phi}
{(\phi - \phi_0)f(\phi)-i\epsilon\sgn(b)}
\nonumber \\
&=& \lim_{\epsilon\to 0} \frac{1}{b}\int_0^{\pi}\frac{d\phi}
{(\phi - \phi_0-i\epsilon\sgn(b))f(\phi)}
\eea
since $f(\phi_0)$ is positive.  Our standard formula Eq.~(\ref{celeb})
can be applied to this integral to yield
\be
\Im\left(\lim_{\epsilon\to 0}\frac{1}{b}\int_0^{\pi}\frac{d\phi}
{a/b+\cos\phi-i\epsilon\sgn(b)}\right)
= \frac{1}{b}\pi\sgn(b)\frac{1}{f(\phi_0)}
= \frac{\pi}{\sqrt{b^2-a^2}} \Punkt
\ee
The second integral on the right hand side of Eq.~(\ref{split}) can be
treated in the same fashion and is found to give the same contribution.
QED.

\section{Proof of Eq.~(\ref{FORMEL})}
\label{appformel}
The assertion is
\be
\lim_{\epsilon\to 0}
\int_a^b\frac{dx}{\sqrt{(b-x)(x-a)}(x+c-i\epsilon)} =
\frac{\pi}{\sqrt{|\Delta|}}\left(\overline{\sgn(x+c)}
\Theta(\Delta)+i\Theta(-\Delta)\right)
\ee
with $\Delta=(a+c)(b+c)$.
To prove this, we first apply Eq.~(\ref{celeb}) to obtain
\be
\lim_{\epsilon\to 0}
\int_a^b\frac{dx}{\sqrt{(b-x)(x-a)}(x+c-i\epsilon)} =
\Vp_a^b\frac{dx}{\sqrt{(b-x)(x-a)}(x+c)} +
\frac{i\pi\Theta(-\Delta)}{\sqrt{-\Delta}} \Punkt \label{spalt}
\ee
The $\Theta$ function on the right hand side of Eq.~(\ref{spalt})
emerges from the fact that the singularity occurs only if
\be
a \le -c \le b \quad \Leftrightarrow \quad \Delta \le 0 \Punkt
\ee
The real part of Eq.~(\ref{spalt}) can be looked up in integral tables.
(See e.~g. \cite{GroeHof1}, integral No.~221.7.) For $\Delta>0$,
one obtains
\be
\int \frac{dx}{\sqrt{(b-x)(x-a)}(x+c)} = \frac{1}{\sqrt{\Delta}}
\arcsin\left(\frac{-(c+a)(b-x)+(c+b)(x-a)}{(b-a)|x+c|}\right)
\ee
which gives
\be
\int_a^b\frac{dx}{\sqrt{(b-x)(x-a)}(x+c)} =
\frac{\pi\overline{\sgn(x+c)}}{\sqrt{\Delta}} \Punkt \label{B5}
\ee
For $\Delta<0$, one obtains
\be
\int \frac{dx}{\sqrt{(b-x)(x-a)}(x+c)} =
\frac{1}{\sqrt{-\Delta}}\log
\frac{\left(\sqrt{(c+a)(b-x)}-\sqrt{-(c+b)(x-a)}\right)^2}{|x+c|}
\ee
which gives
\be \Vp_a^b\frac{dx}{\sqrt{(b-x)(x-a)}(x+c)} = 0 \Punkt \label{B7} \ee
Combining Eqs.~(\ref{spalt}), (\ref{B5}) and (\ref{B7}), proves
the assertion.

\end{appendix}

\begin{figure}
\caption[]{Feynman diagram for the irreducible polarization function.}
\label{polar}
\end{figure}

\end{document}